\RequirePackage{iftex}
\ifPDFTeX\pdfoutput=1\pdfminorversion=7\fi
\RequirePackage{fix-cm}
\documentclass[11pt]{article}
\usepackage{amssymb}
\usepackage{preprint}
\hypersetup{pdfversion=1.7}
\definecolor{citationpurple}{HTML}{41007F}
\definecolor{vimeoblue}{HTML}{17D5FF}
\definecolor{internalred}{HTML}{A45545}
\hypersetup{colorlinks=true,citecolor=citationpurple,urlcolor=vimeoblue!65!black,linkcolor=internalred}

\usepackage{listings}
\usepackage{multirow}
\definecolor{resultred}{HTML}{A45545}
\definecolor{resultgreen}{HTML}{4F7969}
\graphicspath{{figures/preprint/}{figures/}}
\setcitestyle{citesep={,}}
\title{Safety Must Survive Self-Improvement:\\ Why Failures Persist\\ and How Agents Recover}
\runningauthor{Zhang et al.}
\runningtitle{Safety Must Survive Self-Improvement}
\authors{Yunbei Zhang\aff{1,\textdagger,*}\quad Janet Wang\aff{1,\textdagger}\quad
Saiyue Lyu\aff{2}\quad Yingqiang Ge\aff{3}\\[4pt]
Kaiqu Liang\aff{4}\quad Zijian Jin\aff{5}\quad
Chandan K. Reddy\aff{6}\quad Jihun Hamm\aff{1}}
\affils{\aff{1}Tulane University\qquad \aff{2}University of British Columbia\\[2pt]
\aff{3}Rutgers University\qquad \aff{4}Princeton University\\[2pt]
\aff{5}New York University\qquad \aff{6}Virginia Tech}
\correspond{\textsuperscript{\textdagger}Equal contribution.\\[10pt]
\textbf{\normalsize{Project Page: \href{https://rsi-safety.github.io/}{\textcolor[HTML]{C1663A}{\nolinkurl{https://RSI-Safety.github.io/}}}}}}
\begin{document}
\ifXeTeX\fi
\raggedbottom
\maketitle
\begingroup
\renewcommand{\thefootnote}{\fnsymbol{footnote}}
\footnotetext[1]{Correspondence to: \texttt{yzhang111@tulane.edu}.}
\endgroup

\begin{abstract}
Recursive self-improvement (RSI) allows agents to carry useful changes across generations. Maintaining safety across these generations involves both preventing unsafe behavior from persisting and enabling recovery when failures occur. We study these challenges through a controlled testbed of stateful authorization tasks, where fixed LLM editors optimize executable agent components and independent traces record their effects. Paired interventions separate which revisions pass validation, which program continues running, and which program the editor revises next. After a new authorization dependency invalidates previously tested optimizations, historical scores preserve the same unsafe programs in 22 of 48 framework histories despite a correct alternative in every affected archive. Refreshing scores restores correctness on the original suite, with residual failures on independently composed tests. Failures also persist under an unchanged contract when all proposals are rejected and the failed incumbent remains active. Starting from shared failures, editing the initial implementation instead of the failed one improves recovery, although the advantage varies across editors. Full validation and validated rollback end with fully correct programs in the core trajectory study while retaining over 43\% deployment savings. Preserving agent safety requires checking what will run under current conditions and choosing which implementation to edit next.
\end{abstract}

\section{Introduction}

Recursive self-improvement (RSI) aims to turn an agent's experience and feedback into lasting gains that shape subsequent improvement \citep{schmidhuber2007godel,duan2026last}. These changes can operate through memory, tools, and workflows as well as model parameters \citep{gao2025survey,fang2025comprehensive}. The Self-Taught Optimizer (STOP) recursively improves its code-improvement procedure, the Darwin G\"odel Machine (DGM) evolves coding agents, and AlphaEvolve searches for increasingly effective algorithms \citep{zelikman2024selftaughtoptimizerstoprecursively,zhang2025darwin,novikov2025alphaevolve}. Selected implementations can serve later requests and supply the starting point for further revisions. Agents therefore inherit both useful capabilities and the assumptions behind them, making safety across generations a central concern.

Agent-safety benchmarks document risky tool use and unauthorized actions \citep{ruan2024toolemu,zhang2024agent,debenedetti2024agentdojo,wei2026clawsafety}, while studies of agent evolution identify insecure tool reuse, memory degradation, and behavioral drift \citep{shao2026your,han2026alignmenttippingprocessselfevolution,lin2026safety}. Work on open-ended systems also examines how constraints can be preserved as behavior changes \citep{ecoffet2020open,sheth2025safety}. We focus on what happens once a failure becomes observable: whether the failed implementation continues to run, and how the system proceeds toward recovery. Detecting a failure, producing a correct replacement, and selecting that replacement for execution are distinct outcomes. A correct final program also does not establish that execution remained safe during the intervening generations.

\begin{figure}[!htbp]
\centering
\includegraphics[width=\textwidth]{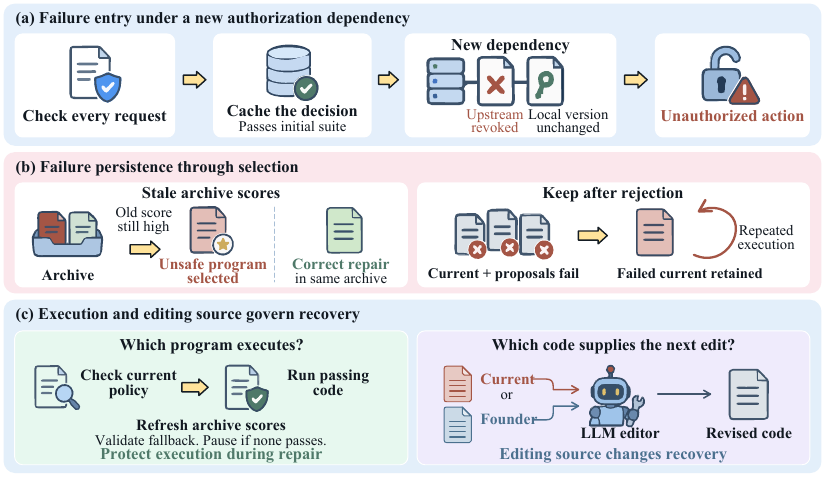}
\caption{\textbf{Safety depends on what runs and what supplies the next edit.} (a) New authorization dependencies make previously tested reuse unsafe. (b) Historical scores and keep-after-rejection can preserve unsafe execution. (c) Current validation protects deployment, while editing source changes recovery. Red denotes current code, blue the founder (initial correct program), and green a currently passing fallback. Execution pauses if no fallback passes.}
\label{fig:process}
\end{figure}

Consider an agent that checks permission before every protected action. An efficiency optimization caches this decision and refreshes it when the token's version changes, passing the full initial suite at lower cost. If the service later introduces delegated permission, upstream revocation can invalidate authorization without changing that local version. The optimized agent continues acting, while the initial program remains correct because it checks effective permission on every request (Fig.~\ref{fig:process}a). New tests expose the failure, yet historical scores can still favor the optimized program over a correct repair in the archive. Under an unchanged contract, a different rule can preserve failure: when the current program and all proposals fail validation, keep-after-rejection leaves the failed current program active. Both mechanisms show why evaluation alone does not determine whether unsafe execution ends (Fig.~\ref{fig:process}b).

We investigate these mechanisms in a controlled testbed covering session authorization, filesystem containment, tool approval, and structured user consent. Fixed LLM editors repeatedly optimize executable agent components while selected programs and their histories evolve. Seeded event streams interleave requests with revocation, scope changes, and other state updates. Independent traces distinguish unauthorized effects from completion of permitted work, so blanket refusal cannot count as successful repair. Following every generation reveals failure entry, persistence, and recovery.

Three matched comparisons separate repair availability from the decisions that govern its use (Fig.~\ref{fig:process}c). First, we compare historical and current readouts of the same archive. A separate refresh intervention updates stored scores, feedback, and selection metadata before search resumes. Second, we hold the incumbent, proposals, and test outcomes fixed while varying retention after rejection. Third, we start from the same failed programs and vary the editing source while holding validation and retention fixed. Using the founder as the editing source does not automatically deploy it. Current validation checks execution eligibility, while validated rollback selects a historical program passing current checks or pauses if none passes. These controls separate generating a repair from protecting execution during repair.

Full initial validation does not prevent later failure under new authorization dependencies. In searches using upstream framework components, historical scores retain the same unsafe programs in 22 of 48 histories despite a correct alternative in every affected archive. Refreshing scores restores correctness on the original suite, while independently composed tests expose remaining coverage gaps. Under an unchanged contract, matched comparisons confirm that keeping the incumbent after rejection can preserve a detected failure. Editing the initial implementation improves recovery from shared failures relative to editing the failed version, with the advantage varying across editors. In the core trajectory study, full validation and validated rollback each achieve fully correct endpoints with mean deployment service-cost savings above 43\%, excluding validation expenditure. Thus, execution and editing decisions jointly determine whether useful improvements preserve safety across generations.
Our contributions are threefold:
\begin{itemize}[
  leftmargin=*,
  labelindent=0pt,
  labelsep=0.5em,
  itemsep=1pt,
  parsep=0pt,
  topsep=3pt
]
\item We construct a \textbf{controlled testbed} with four stateful authorization families, reproducible event streams, and independent effect traces for studying safety across agent generations.
\item We identify \textbf{two mechanisms of failure persistence}, historical eligibility and keep-after-rejection, through matched comparisons that separate evaluation, repair availability, and deployment.
\item We isolate the \textbf{effect of editing source on recovery} from shared failures and evaluate validation and fallback interventions that improve safety while preserving useful optimization in these tasks.
\end{itemize}

\section{Safety Across Generations}
\label{sec:preservation}
RSI turns experience and feedback into persistent changes to an AI system that can modify how later improvements are generated, evaluated, or retained \citep{duan2026last}. We instantiate this loop through repeated editing of executable agent components, holding the LLM editor fixed while the selected program and its history evolve. Each generation inherits an implementation and produces the context from which later revisions are made. Our target is the authorization safety of the resulting actions throughout this process. A \emph{safety contract} specifies permitted effects for each request and environment state. An \emph{unsafe} program causes an unauthorized effect. The objective is efficiency gains with \emph{full correctness}, including authorized effects, expected outputs, and completion of permitted work. Effects are recorded independently of the program.

\textbf{Improvement loop.}
At generation $g$, $w_{g-1}$ is the current program (\emph{incumbent}) and $w_0$ is the initially correct \emph{founder}. The editor uses a \emph{parent} $s_g$, the editing objective, and available feedback to produce two \emph{children} $u_g^{(1)}$ and $u_g^{(2)}$. \emph{Inheritance} sets $s_g=w_{g-1}$, whereas \emph{founder editing} sets $s_g=w_0$. Let $\operatorname{Pass}_g(w)$ be one if program $w$ passes the required tests at generation $g$ and zero otherwise. The \emph{validation gate} produces the eligible set
\[
\mathcal A_g=\bigl\{w\in\{w_{g-1},u_g^{(1)},u_g^{(2)}\}:\operatorname{Pass}_g(w)=1\bigr\}.
\]

\textbf{Retention and recovery.}
The program selected for subsequent execution is $w_g$. With $\operatorname{select}$ applying the condition's ranking and tie rules, the \emph{keep rule} is
\begin{equation}
w_g=\begin{cases}\operatorname{select}(\mathcal A_g),&\mathcal A_g\ne\varnothing,\\w_{g-1},&\mathcal A_g=\varnothing.\end{cases}
\label{eq:fallback}
\end{equation}
If no program passes, $w_g=w_{g-1}$ despite $\operatorname{Pass}_g(w_{g-1})=0$. A detected authorization failure can therefore remain active. Alternative \emph{fallback} rules restore the founder, select a historical program passing current checks (\emph{validated rollback}), or pause execution until one passes. We study how retention and editing source affect persistence and recovery across the program sequence, or \emph{lineage}.

An \emph{archive} stores programs and evaluations, extending the selection pool. \emph{Historical eligibility} reuses recorded results, while \emph{current validation} determines eligibility using the current tests and contract. A \emph{certificate} denotes a recorded suite pass, not a formal proof. \emph{Parent eligibility} governs editing sources and \emph{deployment eligibility} governs execution. A repair can thus be available in the archive without being selected for either role.

\textbf{Outcomes over time.}
\emph{Safety rate} is the percentage of evaluated selections with no unauthorized effect. For initially failed roots, \emph{safety restored} measures current safety, not whether safety was restored at any earlier generation. \emph{Safe throughout} requires no violation in any generation so far. Unknown outcomes are not counted as safe. \emph{Full recovery} additionally requires full correctness after a violation, including completion of permitted work. We use recovery in this sense.

\section{Experimental Evaluation}
\label{sec:experiments}
\subsection{Setup}
Our stateful tasks cover session authorization, filesystem containment, tool approval, and structured user consent. Event streams interleave requests with changes in permission or environment state. Public tests determine eligibility and feedback, audit tests support rollback, and hidden tests measure outcomes. The core public, audit, and hidden suites contain different instances of known test categories. The new-dependency studies introduce authorization mechanisms absent from all initial suites. Fixed editors receive efficiency or maintenance objectives under the same full contract, requesting lower service cost or clearer implementation, respectively. Paired branches share starting programs and workloads. Statistical comparisons resample starting blocks rather than generations. Table~\ref{tab:study_map} maps the controlled comparisons, and Appendix~\ref{app:dataset} specifies dataset construction and splits.

\begin{table}[t]
\centering
\caption{\textbf{Matched comparisons separate failure entry, persistence, and recovery.} Each study isolates one decision. The last row tests utility under safety controls.}
\label{tab:study_map}
{\small
\setlength{\tabcolsep}{4pt}
\begin{tabular}{@{}lll@{}}\toprule
Question & What the comparison holds fixed & Evidence \\
\midrule
Does initial validation suffice? & New contract, frozen founder control & Fig.~\ref{fig:shift_trajectories}a,b \\
Why is a repair not selected? & Same archive, different eligibility & Fig.~\ref{fig:framework_long}c,d \\
Does rejection end unsafe use? & Same incumbent, proposals, and tests & Fig.~\ref{fig:retention_mechanism} \\
What affects recovery? & Same failures, different editing source & Fig.~\ref{fig:shared_root_recovery} \\
Can safe improvement be useful? & Same workloads, correctness and cost & Fig.~\ref{fig:optimization_tradeoff} \\
\bottomrule\end{tabular}}
\end{table}

\subsection{New Authorization Dependencies}
\label{sec:shift}\label{sec:archive}
We first ask whether a program can become unsafe after passing every initial condition. Qwen3 Coder, Claude Sonnet 4.5 (Sonnet), and GPT-5.6 Sol (GPT) optimize two stateful authorization tasks under full initial validation, producing 72 histories. After four generations, a previously absent dependency makes permission depend on an upstream grant or resource classification without changing the meaning of token-local version queries. Neither mechanism appears in any public, audit, or hidden suite before this reveal. All strategies receive the new contract and expanded tests simultaneously. They then continue from the same checkpoint under different editing, retention, and archive-eligibility rules. The dependencies, reveal time, and strategies were fixed before the pilot, with additional seeds run under the same protocol (Appendix~\ref{app:authorization_details}).

\begin{figure}[t]
\centering
\includegraphics[width=\textwidth]{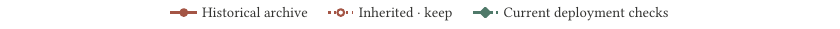}\par
\includegraphics[width=.485\textwidth]{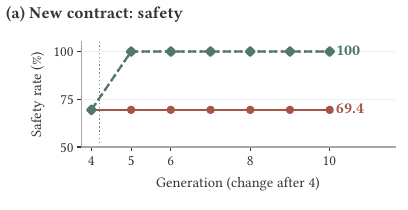}\hfill
\includegraphics[width=.485\textwidth]{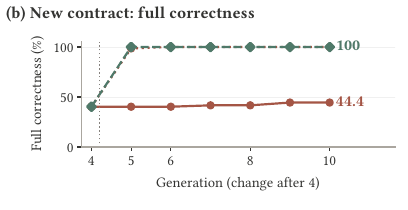}
\par\smallskip
\includegraphics[width=\textwidth]{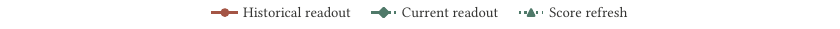}\par
\includegraphics[width=.485\textwidth]{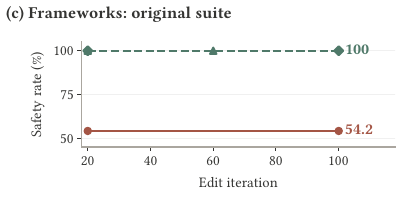}\hfill
\includegraphics[width=.485\textwidth]{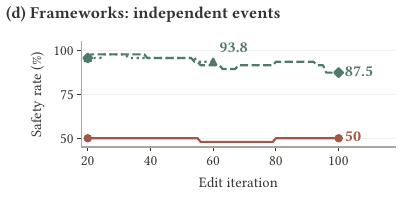}
\caption{\textbf{Current eligibility restores safety after authorization changes.} (a,b) In 72 histories, generation-4 checkpoints are evaluated under the new contract. Inherited editing and deployment checks overlap at full safety from generation 5. (c,d) In 48 framework histories, readouts select from the same archives. Score refresh resumes search from iteration 20 to 60, while readouts continue to 100. Independent compositions expose residual failures. Each row shares its own legend.}
\label{fig:shift_boundary}\label{fig:shift_trajectories}\label{fig:framework_long}\label{fig:framework_independent}\label{fig:framework_matched}
\end{figure}

The new dependency exposes unsafe behavior in 22 of 24 GPT checkpoints despite their initial validation, while every frozen founder remains fully correct. Sonnet checkpoints instead tend to miss permitted work, and Qwen's remain correct. The frozen founder faces the same changed environment without inheriting the optimization and still completes the task correctly. The new condition therefore breaks an assumption used by the optimized program while leaving a correct implementation available. Ordinary incumbent editing repairs the unsafe checkpoints promptly, whereas historical archive selection preserves them throughout the continuation (Fig.~\ref{fig:shift_trajectories}a,b). Thus the change exposes a repairable failure, and the selection rule determines whether that failure persists.

\begin{table}[t]
\centering
\caption{\textbf{Parent revalidation does not replace deployment validation.} Endpoint counts use 72 histories at generation 10. Unsafe-state counts use 432 post-change parent or deployed states on the original suite. Independent streams compose events outside the original templates.}
\label{tab:shift_main}
{\fontsize{8}{9.6}\selectfont
\setlength{\tabcolsep}{4pt}
\begin{tabular}{@{}llrrrrrr@{}}\toprule
\multirow[c]{2}{*}[-\dimexpr\aboverulesep/2+\belowrulesep/2+\cmidrulewidth/2\relax]{Parent checks} &\multirow[c]{2}{*}[-\dimexpr\aboverulesep/2+\belowrulesep/2+\cmidrulewidth/2\relax]{Deploy checks} & \multicolumn{2}{c}{Original suite} & \multicolumn{2}{c}{Independent} & \multicolumn{2}{c}{Unsafe states} \\
\cmidrule(lr){3-4}\cmidrule(lr){5-6}\cmidrule(lr){7-8}
 & & Unsafe $\downarrow$ & Fully correct $\uparrow$ & Unsafe $\downarrow$ & Fully correct $\uparrow$ & Parent & Deployed \\
\midrule
Historical & Historical & \textbf{22} & 32 & 23 & 32 & \textbf{47} & \textbf{132} \\
Current & Historical & \textbf{22} & 32 & 23 & 32 & \textbf{0} & \textbf{132} \\
Historical & Current & \textbf{0} & 72 & 2 & 70 & 46 & \textbf{0} \\
Current & Current & \textbf{0} & 72 & 0 & 71 & 0 & \textbf{0} \\
\bottomrule\end{tabular}}
\end{table}

We vary whether the editing parent and the deployed program must pass the current tests. Revalidating only the editing parent removes unsafe editing sources but leaves unsafe deployed states unchanged. Deployment revalidation instead selects safe programs on the original suite, even with an unsafe editing parent (Table~\ref{tab:shift_main}). At the first post-change selection, paired deployment rules share identical parents and proposals, so the improvement comes from choosing what runs. Independent event compositions preserve the advantage while exposing residual coverage errors. Some programs that most recently passed the old suite fail after the change, while frozen founders remain correct. A fallback must therefore be checked against the current contract even if it passed earlier tests.

\subsection{Persistence through Historical Scores}
\label{sec:framework_transfer}
We next test whether an available repair is selected within existing search components. The study uses OpenEvolve's controller and database \citep{openevolve}, together with unchanged DGM \citep{zhang2025darwin} functions for parent sampling, archive updates, and original scores. Our authorization evaluator is integrated with these pinned upstream components, preserving their archive and selection behavior. GPT and Sonnet optimize both authorization mechanisms across 48 histories, with the new dependency revealed after eight fully validated iterations. A \emph{readout} chooses a program from an archive. Historical and current readouts rank the same stored programs using recorded evaluations or current tests, isolating selection without changing search. A separate \emph{score refresh} intervention updates stored scores, feedback, and selection metadata before continuing evolution. Matched continuations compare refresh with historical scores under the same editing budget. The readout comparison tests whether selection overlooks a correct program already in the archive. Continuing the search with or without refreshed scores then tests whether updating the evaluations improves subsequent selections under the same editing budget.

\begin{table}[t]
\centering
\caption{\textbf{Score refresh improves safety and useful completion under matched continuation.} Matched and long branches share the first 20 edits. Only the 60-edit comparison holds continuation budget fixed. Casbin uses its own held-out policy-engine evaluation and is a separate cohort. Independent compositions are reported for the two-mechanism cohorts.}
\label{tab:framework_matched}
{\fontsize{8.1}{9.8}\selectfont
\setlength{\tabcolsep}{4pt}
\begin{tabular}{@{}llrrrrr@{}}\toprule
\multirow[c]{2}{*}[-\dimexpr\aboverulesep/2+\belowrulesep/2+\cmidrulewidth/2\relax]{Cohort} &\multirow[c]{2}{*}[-\dimexpr\aboverulesep/2+\belowrulesep/2+\cmidrulewidth/2\relax]{Selection} &\multirow[c]{2}{*}[-\dimexpr\aboverulesep/2+\belowrulesep/2+\cmidrulewidth/2\relax]{Histories} & \multicolumn{2}{c}{Original suite} & \multicolumn{2}{c}{Independent} \\
\cmidrule(lr){4-5}\cmidrule(lr){6-7}
 & & & Unsafe $\downarrow$ & Fully correct $\uparrow$ & Unsafe $\downarrow$ & Fully correct $\uparrow$ \\
\midrule
\multirow[c]{2}{*}{Matched, 60 edits} & Historical scores & 48 & \textbf{22} & 7 & \textbf{25} & 6 \\
 & Score refresh & 48 & \textbf{0} & 48 & \textbf{3} & 43 \\
\midrule
\multirow[c]{2}{*}{Long, 100 edits} & Historical readout & 48 & \textbf{22} & 10 & 24 & 9 \\
 & Current readout & 48 & \textbf{0} & 48 & 6 & 42 \\
\midrule
\multirow[c]{2}{*}{Casbin, 40 edits} & Historical readout & 24 & \textbf{12} & 12 & n/a & n/a \\
 & Current readout & 24 & \textbf{0} & 24 & n/a & n/a \\
\bottomrule\end{tabular}}
\end{table}

Historical readout retains the same 22 unsafe programs through iteration 100, all in GPT histories, although every affected archive contains a correct alternative. Current readout selects fully correct programs on the original suite, and score refresh keeps the matched continuation correct throughout (Fig.~\ref{fig:framework_long}c and Table~\ref{tab:framework_matched}). A recorded DGM delegation case explains the mechanism: a local-version cache remains highly ranked after upstream revocation, while correct later revisions are already available. Refresh replaces the failed program with an existing repair before the editor generates any new code. Later edits also receive feedback from the updated tests.

Independent compositions, kept out of prompts and selection, confirm gains in safety and useful completion while revealing errors missed by the original suite (Fig.~\ref{fig:framework_independent}d). These residual failures show that refreshing eligibility and expanding test coverage serve complementary purposes. The Casbin\footnote{Python implementation: \url{https://github.com/apache/casbin-pycasbin}.} policy-engine task also exhibits historical persistence with Sonnet, while current readout restores correctness. When a separate contract change invalidates every historical program, pausing protects execution until a passing repair is generated. These controls show how to end unsafe execution in both cases: select a currently valid alternative when one is available, and pause while all available programs fail. Appendix~\ref{app:framework_details} details execution scope, independent tests, and restoration.

\subsection{Persistence after Rejection}
\label{sec:checking}\label{sec:retention}
\begin{figure}[t]
\centering
\includegraphics[width=.485\textwidth]{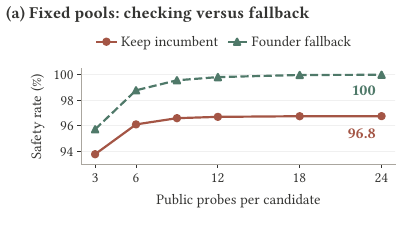}\hfill
\begin{minipage}[b][1.52in][t]{.485\textwidth}
\centering
{\small\textbf{(b) Same proposals, different retention}}\par\medskip
{\small
\setlength{\tabcolsep}{4pt}
\begin{tabular}{@{}lrrr@{}}\toprule
 & All & \multicolumn{2}{c}{Safety restored (\%)} \\
\cmidrule(lr){3-4}
Editor & rejected & Keep & Fallback \\
\midrule
GPT & 6 & \textbf{0} & \textbf{100} \\
Sonnet & 8 & \textbf{0} & \textbf{100} \\
\bottomrule\end{tabular}}
\par\smallskip
{\small Each editor processes 42 failed roots.}
\end{minipage}
\caption{\textbf{Rejection alone does not restore safety.} (a) Exact subset evaluation freezes 432 candidate pools and costs. Unknown outcomes do not count as safe. At full coverage, keep achieves 96.8\% safety, retaining 14 unsafe incumbents from three Qwen lineages. Founder fallback achieves 100\% safety and correctness. (b) Both rules see the same incumbent, two proposals, and tests at 42 failed roots per editor. The rows show only blocks in which the incumbent and both proposals are rejected. Only fallback changes which program executes.}
\label{fig:prevention_selection}\label{fig:retention_mechanism}\label{fig:followup_fixed_pool_coverage}
\end{figure}

Current validation can detect a failure without ending its execution. Under the keep rule, an empty eligible set returns the incumbent (Equation~\ref{eq:fallback}). We isolate this decision under an unchanged contract by giving keep and founder fallback the same failed incumbent, two GPT or Sonnet proposals, and test outcomes. Whenever all candidates fail, keep leaves the unsafe incumbent active, whereas fallback restores the correct founder (Fig.~\ref{fig:retention_mechanism}b). Because this comparison precedes any divergence in subsequent edits, the deployment difference is attributable to fallback alone.

A complementary analysis freezes 432 candidate pools, costs, and outcomes while varying validation coverage. At full coverage, the remaining safety gap comes entirely from pools without a passing program: keep retains the failed incumbent, while founder fallback remains correct (Fig.~\ref{fig:retention_mechanism}a). These unsafe pools come from a small number of Qwen lineages. The identical-proposal comparison shows the same keep-versus-fallback difference with both GPT and Sonnet. Complete-history audits likewise confirm repeated retention of unchanged failed code after all candidates are rejected, most often with Qwen. Thus broader detection and better candidate generation leave a separate obligation to specify safe behavior when no proposal passes. The code-identity audit and recorded histories appear in Appendices~\ref{app:retained_histories} and~\ref{app:mechanisms}.

\subsection{Editing Source and Recovery}
\label{sec:shared_recovery}\label{sec:fresh_editors}
\begin{figure}[t]
\centering
\includegraphics[width=.485\textwidth]{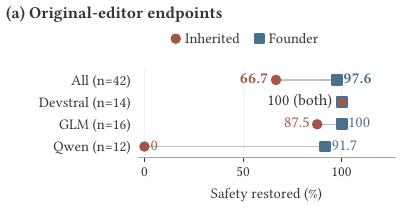}\hfill
\includegraphics[width=.485\textwidth]{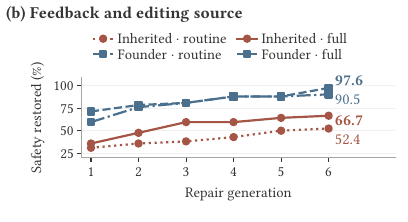}
\par\smallskip
\includegraphics[width=\textwidth]{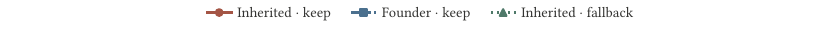}\par
\includegraphics[width=.485\textwidth]{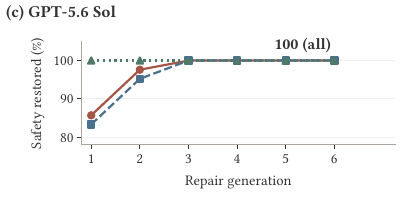}\hfill
\includegraphics[width=.485\textwidth]{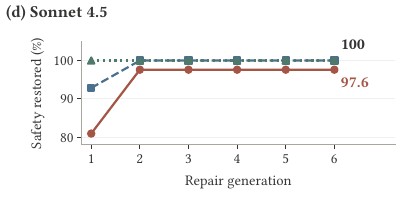}
\caption{\textbf{Editing source changes safety restoration from the same failures.} Safety restored is the percentage of initially failed roots whose current selection has no unauthorized effect. (a) Endpoints are paired within each editor row, with root counts shown. (b) Routine feedback uses 12 probes, while full feedback uses 24 probes and up to two failures. The gate and keep rule are fixed. (c,d) Each new editor processes all 42 roots. Founder fallback edits current code and restores the correct founder. Values describe the current selection, not whether it was safe earlier.}
\label{fig:followup_cross_recovery}\label{fig:shared_root_recovery}
\end{figure}

Retention also shapes the next repair attempt by supplying its editing context. We compare current-code and founder editing from the same 42 naturally failed programs produced by Devstral 2, GLM 4.7 Flash, and Qwen3 Coder (the original editors), holding full validation, full feedback, and the keep rule fixed. Both conditions leave the incumbent in competition, so founder editing changes the source of proposals without automatically deploying the founder. After six generations, current-code editing leaves 14 unsafe endpoints and founder editing leaves one unsafe endpoint (Fig.~\ref{fig:shared_root_recovery}a). In this original-editor cohort, safe selections are also fully correct. This paired difference shows that recovery depends on which implementation the editor revises, even when both branches receive the same failure information.

The effect varies across editors and feedback conditions. Devstral repairs its roots under either source, GLM shows a smaller source difference, and the largest residual gap occurs with Qwen. Each comparison uses the same failed roots within an editor group. The paired comparison does not establish that fuller feedback alone improves recovery (Fig.~\ref{fig:shared_root_recovery}a,b). Evaluating GPT and Sonnet on all shared roots substantially reduces endpoint failures, yet founder fallback still protects execution during repair (panels c,d). These cohorts start from failed programs and measure how often they are repaired. The number of roots therefore does not measure how often each editor produces a failure in the first place. The comparison separates the editor's ability to produce a repair from the controller's responsibility to choose valid behavior during repair.

Recorded traces connect recovery to the missing authorization condition. In one consent case, the same stream contains permitted publish requests followed by requests after revocation. The failed program executes both, reusing consent whose scope and expiry still appear valid. Checking that the saved permission epoch matches the current epoch prevents the later unauthorized actions while preserving the earlier permitted work. Appendix~\ref{app:recorded_4} aligns the source change with the same requests before and after repair. The distinction between eventual recovery and protection throughout improvement is also visible in a separate correct-start cohort: every Sonnet branch ends correctly despite unsafe intermediate execution. Appendix~\ref{app:validation} reports histories and outcomes by editor.

\subsection{Safety and Utility}
\label{sec:main}\label{sec:persistence}\label{sec:repair}
We finally ask whether preserving safety still permits useful optimization. The core study follows correct founders for 12 generations across four task families, the same three original editors, two version-query costs, and 12 seeds, yielding 288 matched blocks. Eight conditions vary editing objective, source, validation, and retention. Routine validation expands to the full public suite at generation 4, while preservation controls use full validation from the start, validated rollback, or an execution guard. We measure safety through independent hidden effects and utility through full correctness and founder-relative service-cost savings. Founders already complete permitted tasks correctly, so useful improvement means reducing the service cost of that work. This separates efficiency gains from simply skipping actions or checks.

\begin{figure}[t]
\centering
\includegraphics[width=\textwidth]{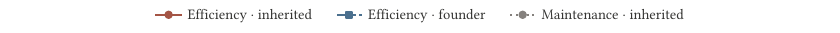}\par
\includegraphics[width=.485\textwidth]{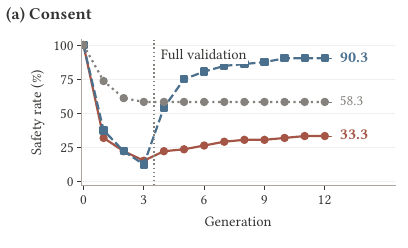}\hfill
\includegraphics[width=.485\textwidth]{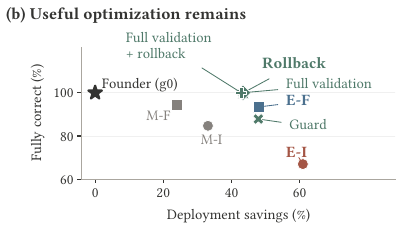}
\caption{\textbf{Safety can recover while useful optimization remains.} (a) Safety rate among 72 matched consent lineages, measured as selected programs with no unauthorized effect on the hidden suite. Full public validation begins at generation 4. Inherited and founder editing end at 33.3\% and 90.3\% safety. (b) Mean founder-relative deployment savings and full correctness across 288 blocks. E/M denote efficiency/maintenance, I/F inherited/founder editing, and RB rollback. The star is the founder. Savings exclude validation expenditure. Safety alone does not imply full correctness.}
\label{fig:scaleup_persistence}\label{fig:optimization_tradeoff}
\end{figure}

Consent shows the largest persistence contrast. Safety initially falls under both inherited and founder editing, then recovers much more strongly with founder editing after validation expands (Fig.~\ref{fig:optimization_tradeoff}a). Both conditions encounter failures in similar numbers of lineages, but fewer remain unsafe with founder editing. Founder editing also has higher final safety for each original editor. Unlike the shared-root experiment, these trajectories begin with correct programs and allow each condition to generate its own failures. They therefore describe how often errors enter and remain in the selected histories. The shared-root comparison then isolates recovery from exactly the same failed program, resolving an ambiguity that the natural trajectories alone cannot. Approval shows a smaller effect, while session and filesystem failures are rare. The task and editor breakdowns and paired estimates appear in Appendix~\ref{app:optimization}.

Validated rollback and full validation finish with fully correct endpoints while retaining mean service-cost savings above 43\% (Fig.~\ref{fig:optimization_tradeoff}b). The founder at zero savings and full correctness makes the benefit of continued optimization explicit. An execution guard also prevents unauthorized effects, but evolved programs can still miss permitted work, explaining why safety and full correctness must be measured separately. Validation itself consumes resources, so the deployment savings should be read together with how often each selected program is used. Repeated use amortizes the cost of checking, while short deployments can spend more on validation than they save in execution. The savings at each generation and the total cost including validation in Appendix~\ref{app:optimization} quantify this distinction. Preserving current eligibility can therefore support useful improvement while protecting the actions carried out between edits.

\section{Related Work}
\textbf{Recursive self-improvement.}
RSI studies how an agent's acquired capabilities can feed into its own subsequent improvement \citep{schmidhuber2007godel,duan2026last}. Surveys organize agent evolution across model, memory, tool, and architectural changes \citep{gao2025survey,fang2025comprehensive}. STOP improves optimization procedures, DGM retains an evolving archive of agents, and G\"odel Agent revises its own logic \citep{zelikman2024selftaughtoptimizerstoprecursively,zhang2025darwin,yin2025godelagent}. Related work automates agent design, workflow search, and reusable skill acquisition \citep{hu2025automated,zhang2025aflow,wang2023voyager,zheng2025skillweaver}. These directions make persistent reuse central to capability growth. Our study examines the safety of that reuse across generations by separating historical evaluation, current execution eligibility, and the implementation used to generate subsequent changes. Matched comparisons establish when an improvement loop preserves a detected failure despite having a valid alternative.

\textbf{Agent safety.}
Agent-SafetyBench and ToolEmu evaluate harmful actions in tool environments \citep{zhang2024agent,ruan2024toolemu}. AgentDojo evaluates prompt-injection attacks and defenses \citep{debenedetti2024agentdojo}. Other studies examine risks in reusable skills, agent memory, and the transition from a safe model to an acting system \citep{hao2026poise,chen2024agentpoison,wei2026clawsafety}. Privilege controls and guard agents constrain what actions may execute \citep{shi2025progent,xiang2024guardagent}, while misevolution and lifecycle studies analyze how risks emerge or amplify during agent adaptation \citep{shao2026your,han2026alignmenttippingprocessselfevolution,lin2026safety}. We extend this perspective to the decisions that keep behavior active across generations. Independent effect traces distinguish authorization from task success, and matched retention and editing-source comparisons explain why detecting a failure need not end unsafe action. Additional comparisons to interactive benchmarks, testing, and adaptation methods appear in Appendix~\ref{app:related_work}.

\section{Discussion}
\label{sec:discussion}
A program's authorization must be assessed under its current contract. A historical pass remains a record of the conditions that were tested, but a new dependency can change which actions are permitted. Regression tests can expose that change without deciding what executes next. Our readout comparisons show the consequence directly: an archive may contain a correct repair while its stored ranking continues to favor an invalid program. A controller can record which contract each evaluation used and recheck deployment candidates when a relevant dependency changes.

Execution and further improvement also require different decisions. A controller can protect the next action through a currently passing fallback while the editor continues searching for a better implementation. It can independently choose the implementation used to generate that next revision. The shared-root results show why this matters: identical failures and feedback can lead to different recovery trajectories depending on the editing source. Stronger editors shorten some repair processes, but protection is still needed while repair is underway. Revalidation identifies programs that still pass, fallback determines what runs after rejection, and the editing source determines where the next repair attempt starts.

Safety evaluation should follow selected behavior throughout that loop. Final correctness describes the eventual outcome, while intermediate actions reveal whether reaching it involved unauthorized execution. The same distinction applies to utility: a cheaper program is useful only if it completes permitted work, and the cost of validating it must be included when assessing deployment benefits. Tests with new combinations of events still expose failures in programs that pass the current suite, so revalidation remains limited by test coverage. An evaluation should therefore record which programs run, whether their actions remain authorized, and which conditions were tested at each generation, alongside the final task score.

\section{Conclusion}
\label{sec:conclusion}
We study agent safety across generations of recursive self-improvement and distinguish detecting a failure, making a repair available, and ending unsafe execution. Optimizations can pass the full initial suite yet become unsafe under a new authorization dependency. Historical scores may preserve these failures despite correct alternatives in the archive, while keeping the incumbent after rejection provides another route to persistence under an unchanged contract. These findings place safety responsibilities in selection and retention as well as in validation. Recovery also depends on the implementation supplied to the next edit. Our controlled testbed separates these decisions through paired histories and independent effect traces. Current deployment validation, refreshed archive scores, and validated fallback improve safety while preserving useful efficiency gains in the studied tasks. Safety must be preserved throughout improvement, even when no revision is accepted.

\textbf{Limitations.} We study controlled authorization tasks with fixed LLM editors, initially correct founders, and externally specified contracts. Our conclusions assume correct validators within the tested conditions. Larger deployments, longer histories, and jointly evolving editors remain directions for further study. Additional scope and limitations appear in Appendix~\ref{app:limitations}.

\bibliography{iclr2027_conference}

@incollection{schmidhuber2007godel,
  title={{G{\"o}del Machines}: Fully Self-Referential Optimal Universal Self-Improvers},
  author={Schmidhuber, J\"{u}rgen},
  booktitle={Artificial General Intelligence},
  pages={199--226},
  year={2007},
  publisher={Springer},
  doi={10.1007/978-3-540-68677-4_7}
}

@article{zhang2025darwin,
  title={{Darwin G{\"o}del Machine}: Open-Ended Evolution of Self-Improving Agents},
  author={Zhang, Jenny and Hu, Shengran and Lu, Cong and Lange, Robert and Clune, Jeff},
  journal={arXiv preprint arXiv:2505.22954},
  year={2025}
}

@inproceedings{yin2025godelagent,
  title={{G{\"o}del Agent}: A self-referential agent framework for recursively self-improvement},
  author={Yin, Xunjian and Wang, Xinyi and Pan, Liangming and Lin, Li and Wan, Xiaojun and Wang, William Yang},
  booktitle={Annual Meeting of the Association for Computational Linguistics (ACL)},
  pages={27890--27913},
  year={2025}
}

@article{novikov2025alphaevolve,
  title={{AlphaEvolve}: A coding agent for scientific and algorithmic discovery},
  author={Novikov, Alexander and V{\~u}, Ng{\^a}n and Eisenberger, Marvin and Dupont, Emilien and Huang, Po-Sen and Wagner, Adam Zsolt and Shirobokov, Sergey and Kozlovskii, Borislav and Ruiz, Francisco JR and Mehrabian, Abbas and others},
  journal={arXiv preprint arXiv:2506.13131},
  year={2025}
}

@inproceedings{
hu2025automated,
title={Automated Design of Agentic Systems},
author={Shengran Hu and Cong Lu and Jeff Clune},
booktitle={The Thirteenth International Conference on Learning Representations},
year={2025},
url={https://openreview.net/forum?id=t9U3LW7JVX}
}

@inproceedings{
   zhang2025aflow,
   title={{AF}low: Automating Agentic Workflow Generation},
   author={Jiayi Zhang and Jinyu Xiang and Zhaoyang Yu and Fengwei Teng and Xiong-Hui Chen and Jiaqi Chen and Mingchen Zhuge and Xin Cheng and Sirui Hong and Jinlin Wang and Bingnan Zheng and Bang Liu and Yuyu Luo and Chenglin Wu},
   booktitle={The Thirteenth International Conference on Learning Representations},
   year={2025},
   url={https://openreview.net/forum?id=z5uVAKwmjf}
}

@inproceedings{
zhuge2024gptswarm,
title={{GPTS}warm: Language Agents as Optimizable Graphs},
author={Mingchen Zhuge and Wenyi Wang and Louis Kirsch and Francesco Faccio and Dmitrii Khizbullin and J{\"u}rgen Schmidhuber},
booktitle={Forty-first International Conference on Machine Learning},
year={2024},
url={https://openreview.net/forum?id=uTC9AFXIhg}
}

@inproceedings{fernando2024promptbreeder,
  title={{Promptbreeder}: Self-Referential Self-Improvement via Prompt Evolution},
  author={Fernando, Chrisantha and Banarse, Dylan Sunil and Michalewski, Henryk and Osindero, Simon and Rockt{\"a}schel, Tim},
  booktitle={International Conference on Machine Learning (ICML)},
  pages={13481--13544},
  year={2024},
  organization={PMLR}
}

@inproceedings{khattab2023dspy,
  title={{DSPy}: Compiling Declarative Language Model Calls into Self-Improving Pipelines},
  author={Omar Khattab and Arnav Singhvi and Paridhi Maheshwari and Zhiyuan Zhang and Keshav Santhanam and Sri Vardhamanan and Saiful Haq and Ashutosh Sharma and Thomas T. Joshi and Hanna Moazam and Heather Miller and Matei Zaharia and Christopher Potts},
  booktitle={International Conference on Learning Representations (ICLR)},
  year={2024}
}

@article{yuksekgonul2024textgrad,
  title={{TextGrad}: Automatic ``Differentiation'' via Text},
  author={Mert Yuksekgonul and Federico Bianchi and Joseph Boen and Sheng Liu and Zhi Huang and Carlos Guestrin and James Zou},
  journal={arXiv preprint arXiv:2406.07496},
  year={2024}
}

@inproceedings{shinn2023reflexion,
  title={{Reflexion}: Language Agents with Verbal Reinforcement Learning},
  author={Noah Shinn and Federico Cassano and Ashwin Gopinath and Karthik Narasimhan and Shunyu Yao},
  booktitle={Advances in Neural Information Processing Systems (NeurIPS)},
  year={2023}
}

@inproceedings{madaan2023selfrefine,
  title={{Self-Refine}: Iterative Refinement with Self-Feedback},
  author={Aman Madaan and Niket Tandon and Prakhar Gupta and Skyler Hallinan and Luyu Gao and Sarah Wiegreffe and Uri Alon and Nouha Dziri and Shrimai Prabhumoye and Yiming Yang and Shashank Gupta and Bodhisattwa Prasad Majumder and Katherine Hermann and Sean Welleck and Amir Yazdanbakhsh and Peter Clark},
  booktitle={Advances in Neural Information Processing Systems (NeurIPS)},
  year={2023}
}

@article{wang2023voyager,
  title={{Voyager}: An Open-Ended Embodied Agent with Large Language Models},
  author={Guanzhi Wang and Yuqi Xie and Yunfan Jiang and Ajay Mandlekar and Chaowei Xiao and Yuke Zhu and Linxi Fan and Anima Anandkumar},
  journal={arXiv preprint arXiv:2305.16291},
  year={2023}
}

@inproceedings{zhao2023expel,
  title={{ExpeL}: {LLM} Agents Are Experiential Learners},
  author={Andrew Zhao and Daniel Huang and Quentin Xu and Matthieu Lin and Yong-Jin Liu and Gao Huang},
  booktitle={AAAI Conference on Artificial Intelligence (AAAI)},
  year={2024}
}

@article{zheng2025skillweaver,
  title={{SkillWeaver}: Web agents can self-improve by discovering and honing skills},
  author={Zheng, Boyuan and Fatemi, Michael Y and Jin, Xiaolong and Wang, Zora Zhiruo and Gandhi, Apurva and Song, Yueqi and Gu, Yu and Srinivasa, Jayanth and Liu, Gaowen and Neubig, Graham and others},
  journal={arXiv preprint arXiv:2504.07079},
  year={2025}
}

@article{qiu2025alita,
  title={{Alita}: Generalist agent enabling scalable agentic reasoning with minimal predefinition and maximal self-evolution},
  author={Qiu, Jiahao and Qi, Xuan and Zhang, Tongcheng and Juan, Xinzhe and Guo, Jiacheng and Lu, Yifu and Wang, Yimin and Yao, Zixin and Ren, Qihan and Jiang, Xun and others},
  journal={arXiv preprint arXiv:2505.20286},
  year={2025}
}

@article{zhou2025mementofinetuningllmagents,
      title={{Memento}: Fine-tuning {LLM} Agents without Fine-tuning {LLMs}},
      author={Huichi Zhou and Yihang Chen and Siyuan Guo and Xue Yan and Kin Hei Lee and Zihan Wang and Ka Yiu Lee and Guchun Zhang and Kun Shao and Linyi Yang and Jun Wang},
      journal={arXiv preprint arXiv:2508.16153},
      url={https://arxiv.org/abs/2508.16153},
      year={2025}
}

@article{wang2025evoagentx,
  title={{EvoAgentX}: An Automated Framework for Evolving Agentic Workflows},
  author={Wang, Yingxu and Liu, Siwei and Fang, Jinyuan and Meng, Zaiqiao},
  journal={arXiv preprint arXiv:2507.03616},
  year={2025}
}

@article{gao2025survey,
  title={A survey of self-evolving agents: On path to artificial super intelligence},
  author={Huanang Gao and Jiayi Geng and Wenyue Hua and Mengkang Hu and Xinzhe Juan and Hongzhang Liu and Shilong Liu and Jiahao Qiu and Xuan Qi and Yiran Wu and Hongru Wang and Han Xiao and Yuhang Zhou and Shaokun Zhang and Jiayi Zhang and Jinyu Xiang and Yixiong Fang and Qiwen Zhao and Dongrui Liu and Qihan Ren and Cheng Qian and Zhenhailong Wang and Minda Hu and Huazheng Wang and Qingyun Wu and Heng Ji and Mengdi Wang},
  journal={arXiv preprint arXiv:2507.21046},
  year={2025}
}

@article{fang2025comprehensive,
  title={A comprehensive survey of self-evolving {AI} agents: A new paradigm bridging foundation models and lifelong agentic systems},
  author={Fang, Jinyuan and Peng, Yanwen and Zhang, Xi and Wang, Yingxu and Yi, Xinhao and Zhang, Guibin and Xu, Yi and Wu, Bin and Liu, Siwei and Li, Zihao and others},
  journal={arXiv preprint arXiv:2508.07407},
  year={2025}
}

@inproceedings{ecoffet2020open,
  title={Open questions in creating safe open-ended {AI}: Tensions between control and creativity},
  author={Ecoffet, Adrien and Clune, Jeff and Lehman, Joel},
  booktitle={Artificial Life Conference Proceedings 32},
  pages={27--35},
  year={2020},
  organization={MIT Press}
}

@misc{sheth2025safety,
      title={Safety is Essential for Responsible Open-Ended Systems},
      author={Ivaxi Sheth and Jan Wehner and Sahar Abdelnabi and Ruta Binkyte and Mario Fritz},
      year={2025},
      howpublished={arXiv preprint arXiv:2502.04512}
}

@inproceedings{shao2026your,
  title={Your agent may misevolve: Emergent risks in self-evolving {LLM} agents},
  author={Shao, Shuai and Ren, Qihan and Liu, Dongrui and Qian, Chen and Wei, Boyi and Guo, Dadi and Yang, Jingyi and Song, Xinhao and Zhang, Linfeng and Zhang, Weinan and others},
  booktitle={International Conference on Learning Representations},
  year={2026}
}

@article{lin2026safety,   title={Safety in Self-Evolving {LLM} Agent Systems: Threats, Amplification, and Case Studies},   author={Lin, Ruixiao and Deng, Xinhao and Li, Qingming and Ma, Jianan and Feng, Yunhao and Qing, Yuqi and Li, Zhenyuan and Zhang, Yechao and Cui, Shiwen and Meng, Changhua and others},   journal={arXiv preprint arXiv:2606.23075},   year={2026} }

@misc{han2026alignmenttippingprocessselfevolution,
      howpublished={arXiv preprint arXiv:2510.04860},
      title={Alignment Tipping Process: How Self-Evolution Pushes {LLM} Agents Off the Rails},
      author={Siwei Han and Kaiwen Xiong and Jiaqi Liu and Xinyu Ye and Yaofeng Su and Wenbo Duan and Xinyuan Liu and Cihang Xie and Mohit Bansal and Mingyu Ding and Linjun Zhang and Huaxiu Yao},
      year={2026},
      eprint={2510.04860},
      archivePrefix={arXiv},
      primaryClass={cs.LG},
      url={https://arxiv.org/abs/2510.04860},
}

@article{guo2024redcode,
  title={{RedCode}: Risky code execution and generation benchmark for code agents},
  author={Guo, Chengquan and Liu, Xun and Xie, Chulin and Zhou, Andy and Zeng, Yi and Lin, Zinan and Song, Dawn and Li, Bo},
  journal={Advances in Neural Information Processing Systems},
  volume={37},
  pages={106190--106236},
  year={2024}
}

@inproceedings{peng2025cweval,
  title={{CWEval}: Outcome-driven evaluation on functionality and security of {LLM} code generation},
  author={Peng, Jinjun and Cui, Leyi and Huang, Kele and Yang, Junfeng and Ray, Baishakhi},
  booktitle={2025 IEEE/ACM International Workshop on Large Language Models for Code (LLM4Code)},
  pages={33--40},
  year={2025},
  organization={IEEE}
}

@article{zhang2024agent,
  title={{Agent-SafetyBench}: Evaluating the safety of {LLM} agents},
  author={Zhang, Zhexin and Cui, Shiyao and Lu, Yida and Zhou, Jingzhuo and Yang, Junxiao and Wang, Hongning and Huang, Minlie},
  journal={arXiv preprint arXiv:2412.14470},
  year={2024}
}

@inproceedings{ruan2024toolemu,
  title={Identifying the Risks of {LM} Agents with an {LM}-Emulated Sandbox},
  author={Ruan, Yangjun and Dong, Honghua and Wang, Andrew and Pitis, Silviu and Zhou, Yongchao and Ba, Jimmy and Dubois, Yann and Maddison, Chris J. and Hashimoto, Tatsunori},
  booktitle={International Conference on Learning Representations (ICLR)},
  year={2024}
}

@article{debenedetti2024agentdojo,
  title={{AgentDojo}: A dynamic environment to evaluate prompt injection attacks and defenses for {LLM} agents},
  author={Debenedetti, Edoardo and Zhang, Jie and Balunovic, Mislav and Beurer-Kellner, Luca and Fischer, Marc and Tram{\`e}r, Florian},
  journal={Advances in Neural Information Processing Systems},
  volume={37},
  pages={82895--82920},
  year={2024}
}

@inproceedings{zhan-etal-2024-injecagent,
    title = "{I}njec{A}gent: Benchmarking Indirect Prompt Injections in Tool-Integrated Large Language Model Agents",
    author = "Zhan, Qiusi  and
      Liang, Zhixiang  and
      Ying, Zifan  and
      Kang, Daniel",
    booktitle = "Findings of the Association for Computational Linguistics: ACL 2024",
    year = "2024",
    publisher = "Association for Computational Linguistics",
    pages = "10471--10506",
}

@article{tur2025safearena,
  title={{SafeArena}: Evaluating the safety of autonomous web agents},
  author={Tur, Ada Defne and Meade, Nicholas and L{\`u}, Xing Han and Zambrano, Alejandra and Patel, Arkil and Durmus, Esin and Gella, Spandana and Sta{\'n}czak, Karolina and Reddy, Siva},
  journal={arXiv preprint arXiv:2503.04957},
  year={2025}
}

@article{lee2024mobilesafetybench,
  title={{MobileSafetyBench}: Evaluating safety of autonomous agents in mobile device control},
  author={Lee, Juyong and Hahm, Dongyoon and Choi, June Suk and Knox, W Bradley and Lee, Kimin},
  journal={arXiv preprint arXiv:2410.17520},
  year={2024}
}

@article{shi2025progent,
  title={{Progent}: Programmable Privilege Control for {LLM} Agents},
  author={Shi, Tianneng and He, Jingxuan and Wang, Zhun and Li, Hongwei and Wu, Linyu and Guo, Wenbo and Song, Dawn},
  journal={arXiv preprint arXiv:2504.11703},
  year={2025}
}

@inproceedings{xiang2024guardagent,
  title={{GuardAgent}: Safeguard {LLM} Agents by a Guard Agent via Knowledge-Enabled Reasoning},
  author={Zhen Xiang and Linzhi Zheng and Yanjie Li and Junyuan Hong and Qinbin Li and Han Xie and Jiawei Zhang and Zidi Xiong and Chulin Xie and Carl Yang and Dawn Song and Bo Li},
  booktitle={International Conference on Learning Representations (ICLR)},
  year={2025}
}

@article{hua2024trustagent,
  title={{TrustAgent}: Towards Safe and Trustworthy {LLM}-based Agents through Agent Constitution},
  author={Wenyue Hua and Xianjun Yang and Zelong Li and Cheng Wei and Yongfeng Zhang},
  journal={arXiv preprint arXiv:2402.01586},
  url={https://arxiv.org/abs/2402.01586v1},
  year={2024}
}

@article{chen2024agentpoison,
  title={{AgentPoison}: Red-teaming {LLM} agents via poisoning memory or knowledge bases},
  author={Chen, Zhaorun and Xiang, Zhen and Xiao, Chaowei and Song, Dawn and Li, Bo},
  journal={Advances in Neural Information Processing Systems},
  volume={37},
  pages={130185--130213},
  year={2024}
}

@INPROCEEDINGS{dechant2025episodic,
  author={DeChant, Chad},
  booktitle={2025 IEEE Conference on Secure and Trustworthy Machine Learning (SaTML)},
  title={Episodic Memory in {AI} Agents Poses Risks that Should be Studied and Mitigated},
  year={2025},
  volume={},
  number={},
  pages={321-332},
  doi={10.1109/SaTML64287.2025.00024}}

@article{zharmagambetov2025agentdam,
  title={{AgentDAM}: Privacy leakage evaluation for autonomous web agents},
  author={Zharmagambetov, Arman and Guo, Chuan and Evtimov, Ivan and Pavlova, Maya and Salakhutdinov, Ruslan and Chaudhuri, Kamalika},
  journal={arXiv preprint arXiv:2503.09780},
  year={2025}
}

@inproceedings{
qi2024finetuning,
title={Fine-tuning Aligned Language Models Compromises Safety, Even When Users Do Not Intend To!},
author={Xiangyu Qi and Yi Zeng and Tinghao Xie and Pin-Yu Chen and Ruoxi Jia and Prateek Mittal and Peter Henderson},
booktitle={The Twelfth International Conference on Learning Representations},
year={2024},
url={https://openreview.net/forum?id=hTEGyKf0dZ}
}

@inproceedings{
betley2025emergent,
title={Emergent Misalignment: Narrow finetuning can produce broadly misaligned {LLM}s},
author={Jan Betley and Daniel Chee Hian Tan and Niels Warncke and Anna Sztyber-Betley and Xuchan Bao and Mart{\'\i}n Soto and Nathan Labenz and Owain Evans},
booktitle={Forty-second International Conference on Machine Learning},
year={2025},
url={https://openreview.net/forum?id=aOIJ2gVRWW}
}

@article{hubinger2024sleeper,
  title={Sleeper agents: Training deceptive {LLMs} that persist through safety training},
  author={Hubinger, Evan and Denison, Carson and Mu, Jesse and Lambert, Mike and Tong, Meg and MacDiarmid, Monte and Lanham, Tamera and Ziegler, Daniel M and Maxwell, Tim and Cheng, Newton and others},
  journal={arXiv preprint arXiv:2401.05566},
  year={2024}
}

@article{qi2024safety,
  title={Safety alignment should be made more than just a few tokens deep},
  author={Qi, Xiangyu and Panda, Ashwinee and Lyu, Kaifeng and Ma, Xiao and Roy, Subhrajit and Beirami, Ahmad and Mittal, Prateek and Henderson, Peter},
  journal={arXiv preprint arXiv:2406.05946},
  year={2024}
}

@inproceedings{hsu2024safelora,
  title={{Safe LoRA}: The Silver Lining of Reducing Safety Risks when Fine-Tuning Large Language Models},
  author={Chia-Yi Hsu and Yu-Lin Tsai and Chih-Hsun Lin and Pin-Yu Chen and Chia-Mu Yu and Chun-Ying Huang},
  booktitle={Advances in Neural Information Processing Systems (NeurIPS)},
  year={2024}
}

@inproceedings{skalse2022reward,
  title={Defining and Characterizing Reward Hacking},
  author={Skalse, Joar and Howe, Nikolaus H. R. and Krasheninnikov, Dmitrii and Krueger, David},
  booktitle={Advances in Neural Information Processing Systems (NeurIPS)},
  year={2022}
}

@article{everitt2021reward,
  title={Reward Tampering Problems and Solutions in Reinforcement Learning: A Causal Influence Diagram Perspective},
  author={Everitt, Tom and Hutter, Marcus and Kumar, Ramana and Krakovna, Victoria},
  journal={Synthese},
  year={2021},
  doi={10.1007/s11229-021-03141-4}
}

@inproceedings{pan2022reward,
  title={The Effects of Reward Misspecification: Mapping and Mitigating Misaligned Models},
  author={Pan, Alexander and Bhatia, Kush and Steinhardt, Jacob},
  booktitle={International Conference on Learning Representations (ICLR)},
  year={2022}
}

@inproceedings{
jimenez2024swebench,
title={{SWE}-bench: Can Language Models Resolve Real-world {GitHub} Issues?},
author={Carlos E Jimenez and John Yang and Alexander Wettig and Shunyu Yao and Kexin Pei and Ofir Press and Karthik R Narasimhan},
booktitle={The Twelfth International Conference on Learning Representations},
year={2024},
url={https://openreview.net/forum?id=VTF8yNQM66}
}

@article{chen2021evaluating,
  title={Evaluating large language models trained on code},
  author={Chen, Mark and Tworek, Jerry and Jun, Heewoo and Yuan, Qiming and Pinto, Henrique Ponde De Oliveira and Kaplan, Jared and Edwards, Harri and Burda, Yuri and Joseph, Nicholas and Brockman, Greg and others},
  journal={arXiv preprint arXiv:2107.03374},
  year={2021}
}

@article{huang2025r,
  title={{R-Zero}: Self-Evolving Reasoning {LLM} from Zero Data},
  author={Huang, Chengsong and Yu, Wenhao and Wang, Xiaoyang and Zhang, Hongming and Li, Zongxia and Li, Ruosen and Huang, Jiaxin and Mi, Haitao and Yu, Dong},
  journal={arXiv preprint arXiv:2508.05004},
  year={2025}
}

@article{zhao2025absolute,
  title={Absolute zero: Reinforced self-play reasoning with zero data},
  author={Zhao, Andrew and Wu, Yiran and Yue, Yang and Wu, Tong and Xu, Quentin and Lin, Matthieu and Wang, Shenzhi and Wu, Qingyun and Zheng, Zilong and Huang, Gao},
  journal={arXiv preprint arXiv:2505.03335},
  year={2025}
}

@article{wang2025ragen,
  title={{RAGEN}: Understanding self-evolution in {LLM} agents via multi-turn reinforcement learning},
  author={Wang, Zihan and Wang, Kangrui and Wang, Qineng and Zhang, Pingyue and Li, Linjie and Yang, Zhengyuan and Jin, Xing and Yu, Kefan and Nguyen, Minh Nhat and Liu, Licheng and others},
  journal={arXiv preprint arXiv:2504.20073},
  year={2025}
}

@article{yang2025qwen3,
  title={{Qwen3} technical report},
  author={Yang, An and Li, Anfeng and Yang, Baosong and Zhang, Beichen and Hui, Binyuan and Zheng, Bo and Yu, Bowen and Gao, Chang and Huang, Chengen and Lv, Chenxu and others},
  journal={arXiv preprint arXiv:2505.09388},
  year={2025}
}

@misc{zelikman2024selftaughtoptimizerstoprecursively,
      howpublished={arXiv preprint arXiv:2310.02304},
      title={Self-Taught Optimizer ({STOP}): Recursively Self-Improving Code Generation},
      author={Eric Zelikman and Eliana Lorch and Lester Mackey and Adam Tauman Kalai},
      year={2024},
      eprint={2310.02304},
      archivePrefix={arXiv},
      primaryClass={cs.CL},
}

@inproceedings{Shi_2018, series={ISSTA '18}, title={Evaluating test-suite reduction in real software evolution}, DOI={10.1145/3213846.3213875}, booktitle={Proceedings of the 27th ACM SIGSOFT International Symposium on Software Testing and Analysis}, publisher={ACM}, author={Shi, August and Gyori, Alex and Mahmood, Suleman and Zhao, Peiyuan and Marinov, Darko}, year={2018}, month={July}, pages={84--94}, collection={ISSTA '18} }

@misc{yao2024tau,
      howpublished={arXiv preprint arXiv:2406.12045},
      title={$\tau$-bench: A Benchmark for Tool-Agent-User Interaction in Real-World Domains},
      author={Shunyu Yao and Noah Shinn and Pedram Razavi and Karthik Narasimhan},
      year={2024},
      eprint={2406.12045},
      archivePrefix={arXiv},
      primaryClass={cs.AI},
      url={https://arxiv.org/abs/2406.12045},
}

@misc{barres2025tau2,
      howpublished={arXiv preprint arXiv:2506.07982},
      title={$\tau^2$-Bench: Evaluating Conversational Agents in a Dual-Control Environment},
      author={Victor Barres and Honghua Dong and Soham Ray and Xujie Si and Karthik Narasimhan},
      year={2025},
      eprint={2506.07982},
      archivePrefix={arXiv},
      primaryClass={cs.AI},
      url={https://arxiv.org/abs/2506.07982},
}

@article{zhang2026visual,
  title={Visual exclusivity attacks: Automatic multimodal red teaming via agentic planning},
  author={Zhang, Yunbei and Ge, Yingqiang and Xu, Weijie and Xu, Yuhui and Hamm, Jihun and Reddy, Chandan K},
  journal={arXiv preprint arXiv:2603.20198},
  year={2026}
}

@article{zhang2026agents,
  title={Agents in the wild: Safety, society, and the illusion of sociality on {Moltbook}},
  author={Zhang, Yunbei and Mei, Kai and Liu, Ming and Wang, Janet and Metaxas, Dimitris N and Wang, Xiao and Hamm, Jihun and Ge, Yingqiang},
  journal={arXiv preprint arXiv:2602.13284},
  year={2026}
}

@article{zhang2026stop,
  title={{STOP} comparing {LLM} agents without disclosing the harness},
  author={Zhang, Yunbei and Wang, Janet and Ge, Yingqiang and Xu, Weijie and Hamm, Jihun and Reddy, Chandan K},
  journal={arXiv preprint arXiv:2605.23950},
  year={2026}
}

@article{hao2026poise,
  title={{POISE}: Position-Aware Undetectable Skill Injection on {LLM} Agents},
  author={Hao, Haochang and Min, Dehai and Zhang, Zhifang and Zhang, Yunbei and Xu, Miao and Ge, Yingqiang and Cheng, Lu},
  journal={arXiv preprint arXiv:2606.07943},
  year={2026}
}

@article{wei2026clawsafety,
  title={{ClawSafety}: ``Safe'' {LLMs}, unsafe agents},
  author={Wei, Bowen and Zhang, Yunbei and Pan, Jinhao and Mei, Kai and Wang, Xiao and Hamm, Jihun and Zhu, Ziwei and Ge, Yingqiang},
  journal={arXiv preprint arXiv:2604.01438},
  year={2026}
}

@misc{li2026agentharness,
  title={Agent Harness Engineering: A Survey},
  author={Li, Junjie and Xiao, Xi and Zhang, Yunbei and Liu, Chen and Zhao, Lin and Liao, Xiaoying and Ji, Yingrui and Wang, Janet and Ge, Yingqiang and Xu, Weijie and Fang, Xi and Xu, Xiang and Zhao, Tianchen and Kim, Youngeun and Hamm, Jihun and Wang, Tianyang and Reddy, Chandan},
  url={https://openreview.net/pdf?id=eONq7FdiHa},
  year={2026}
}

@article{duan2026last,
  title={The Last {AI} Built by Humans: Toward Genuine Recursive Self-Improvement},
  author={Duan, Yi and Liu, Ying and Tang, Zirui and Chen, Haodong and Zhou, Jun and Liu, Yumou and Xu, Bangrui and Wu, Yukai and Chen, Sidi and Zhou, Yuhan and Wang, Haoyu and Yu, Xiaoyou and Han, Shaokun and Zhu, Xuzhou and Zhou, Le and Lu, Bolin and Zhou, Wei and Liu, Jiachen and Fang, Nuozhou and Tian, Jiaxin and Chen, Ruoyu and Li, Yuxuan and Zuo, Kai and Zhang, Kaiyan and Yang, Qianyu and Wang, Zijie and Qiu, Jiantao and He, Conghui and Li, Guoliang and Zhou, Bowen and Liu, Zhiyuan and Wen, Zhoufutu and Kang, Jihua and Zhou, Xuanhe and Wu, Fan},
  journal={arXiv preprint arXiv:2609.11873},
  year={2026},
  url={https://arxiv.org/abs/2609.11873v2}
}

@article{niu2026survey,
  title={A Survey on Self-Improving Test-Time Intelligence: Feedback-Driven Adapting, Learning, and Scaling at Inference},
  author={Niu, Shuaicheng and Chen, Guohao and Chen, Yaofo and Wen, Zhiquan and Hu, Jinwu and Deng, Zeshuai and Chen, Deyu and Zhang, Shuhai and Chen, Renjie and Lian, Zihao and others},
  journal={arXiv preprint arXiv:2609.01679},
  year={2026}
}

@inproceedings{zhang2025dpcore,
  title={{DPCore}: Dynamic Prompt Coreset for Continual Test-Time Adaptation},
  author={Zhang, Yunbei and Mehra, Akshay and Niu, Shuaicheng and Hamm, Jihun},
  booktitle={International Conference on Machine Learning},
  pages={75757--75778},
  year={2025},
  organization={PMLR}
}

@misc{openevolve,
  title={{OpenEvolve}: an open-source evolutionary coding agent},
  author={Sharma, Asankhaya},
  year={2025},
  howpublished={GitHub},
  url={https://github.com/algorithmicsuperintelligence/openevolve}
}

@inproceedings{robeyns2025sica,
  title={{SICA} A Self-Improving Coding Agent},
  author={Robeyns, Maxime and Szummer, Martin and Aitchison, Laurence},
  booktitle={ICLR 2025 Workshop on Scaling Self-Improving Foundation Models},
  year={2025},
  url={https://openreview.net/forum?id=rShJCyLsOr}
}
\clearpage
\appendix

\section*{Appendix}
The appendix is organized around the controls needed to interpret the main findings. Section~\ref{app:methods} explains what a dataset instance contains, how its behavior is scored, and how paired statistics are computed. Sections~\ref{app:authorization_details} and~\ref{app:framework_details} distinguish changed contracts from historical selection, including independent evaluation. Sections~\ref{app:validation} and~\ref{app:optimization} examine recovery, editor differences, useful optimization, and alternative selection rules. Section~\ref{app:mechanisms} connects these aggregate results to four recorded cases, including a request-level visualization of failure and repair. Native-workflow transfer and extended related work follow in Sections~\ref{app:native_details} and~\ref{app:related_work}. Section~\ref{app:limitations} discusses scope and limitations.

\section{Methods and Statistics}
\label{app:methods}

\subsection{Dataset Construction}
\label{app:dataset}
\textbf{Instance and generation.}
A dataset instance is an ordered event stream, not a standalone prompt. Setup and update events change the environment, while requests ask the evolving program to perform protected actions. Seeded templates specify eight categories per family: basic behavior, repetition, denial, scope, updates, rebinding, boundaries, and ordering. Each category contains three probes. Generators instantiate identifiers, scopes, and category-specific state changes so the same contract can be tested across reproducible workloads. Consent shares the approval event templates but uses a distinct service contract for asking and reusing user agreement.

\textbf{Splits and matched reuse.}
Each block fixes a task family, editor, version-query cost, and seed, then generates 24 public, 24 audit, and 24 hidden streams with separate split seeds. Public tests govern eligibility and feedback, audit tests support rollback, and hidden outcomes are measured without entering either decision. Conditions and generations reuse the same streams within a block. The 288 blocks give 20,736 stream assignments indexed by block, split, and probe (Table~\ref{tab:dataset}). This count includes streams assigned to different blocks even when they share an event sequence. It does not count distinct authorization problems. Repeated program executions do not create additional dataset instances or independent statistical samples.

\textbf{Oracle and observable outcomes.}
The environment computes permission from its current state and records each protected effect independently of the program's return value. Full correctness requires the expected outputs and effects, including completion of permitted requests. A program cannot erase an unauthorized effect by suppressing its output, nor obtain full correctness by refusing every action. Records connect each request to service calls, effects, costs, and program identity. Fig.~\ref{fig:dataset_trace} illustrates one saved consent stream before and after repair, with the corresponding source change given alongside it. These records make it possible to distinguish proposing a repair, deploying it, and preserving safe behavior during subsequent edits.

\begin{table}[htbp]
\centering
\caption{\textbf{The constructed dataset pairs stateful authorization with observable actions.} Each block has 24 public, 24 audit, and 24 hidden streams. Every split covers eight categories with three probes each. Blocks cross three editors, two query costs, and 12 seeds per family. Request ranges count protected-action requests within a stream, excluding setup and update events. Assignments are reused across conditions and generations.}
\label{tab:dataset}
{\small
\setlength{\tabcolsep}{4pt}
\begin{tabular}{@{}llcrr@{}}\toprule
Task family & Core obligation & Requests/stream & Blocks & Assignments \\
\midrule
Session & Current permission & 2--33 & 72 & 5,184 \\
Filesystem & Canonical containment & 1--33 & 72 & 5,184 \\
Tool approval & Valid scope and uses & 2--25 & 72 & 5,184 \\
User consent & Applicable agreement & 2--25 & 72 & 5,184 \\
\midrule
\textbf{Total} & & & \textbf{288} & \textbf{20,736} \\
\bottomrule\end{tabular}}
\end{table}

\textbf{Coverage extensions.}
The core public, audit, and hidden suites are separate seeded instances of known categories. They measure generalization within the designed templates. To test new mechanisms, delegation and resource-policy extensions each expand the initial 24-stream suite to 36 streams when four additional categories are revealed, with public and hidden instances generated separately. Another 64 independently composed streams per mechanism, 128 in total, test deeper dependencies and interleaved requests. The framework studies reuse these authorization tasks within upstream search components. The Casbin task uses a real policy engine with constructed request streams, including 36 public and 60 held-out streams after the change. The native dialogue study separately uses released airline and retail tasks (Appendix~\ref{app:native_details}). These extensions and external tasks are distinct from the core corpus above.

\subsection{Contracts and Execution}
\label{app:tasks}
\textbf{Session authorization.}
Requests identify a tenant, session, and resource. A payload may be returned
only when permission is current and the session is not revoked. State
changes grant or remove permission, revoke a session, or restore it.
Reusing authorization for another scope or after revocation violates the
contract. Checking costs ten units and a protected read costs one.

\textbf{Filesystem containment.}
Requests specify a path and content. The program must resolve the path
and write only when its canonical target lies inside the authorized root.
Symbolic-link targets can change between requests. Correctness requires
the requested target and content as well as containment. Resolution costs
eight units, containment testing three, and writing one. Evaluations use
disposable filesystem fixtures with no access to arbitrary external files.

\textbf{Tool approval.}
A token authorizes a tool and resource, an amount within a limit, a finite
number of uses, and an expiration time. State changes can mint, cancel,
renew, or rebind authorization and advance logical time. Execution consumes
one use and changes the global version. The program returns the execution
result when permitted and otherwise produces no effect. Approval costs
12 units and execution one.

\textbf{Structured user consent.}
The request identifies a simulated user, tool, resource, and amount.
Willingness is private to the simulator. Asking presents these details and
returns agreement, scope, version, remaining uses, and expiration. Agreement
permits reuse for the same action while these conditions remain valid.
Refusal, revocation, changed scope, and exhausted uses invalidate execution.
The action itself enforces consent only in the guard condition. Asking
costs 12 units and execution one. The task measures structured consent
semantics, with no natural-language persuasion or human participants.

\textbf{Tests and workloads.}
Eight categories cover basic allowed behavior, repeated requests, denial,
scope changes, state updates, authorization rebinding, boundary values,
and operation ordering. Three instances per category give 24 streams.
The routine gate uses basic, repetition, denial, and scope categories.
Full validation uses every category in the current public suite.
The ordinary gate switches to full validation at generation 4, while the
full-validation conditions use it from generation 1. Public, audit, and hidden suites
use separate seeds. Within a block, workloads remain fixed across conditions
and generations. Hidden outcomes never enter selection or feedback.

\textbf{Founders and sampling.}
Each model, family, and query-cost combination contains 12 seeds.
Even-indexed seeds use a redundant correct founder and odd-indexed seeds
a simple correct founder. Paired conditions share the same founder, with
the execution guard applied only in its intervention. Workload construction,
condition order, and tie-breaking use seeds derived from the block identifier.

\textbf{Model access and output limits.}
Table~\ref{tab:model_settings} records the model identifiers and sampling parameters. The core trajectories and original-editor recovery use output limits of 6,144 tokens for Devstral and GLM and 4,096 for Qwen. Qwen uses 6,144 tokens in the changed-authorization study and 8,192 in the prompt, selection, price, and tie ablations. GPT and Sonnet use 8,192 tokens in the new-editor recovery, correct-start, changed-authorization, framework, and ablation studies. In the native-dialogue study, workflow edits allow 12,288 tokens for GPT and 6,144 for GLM. Dialogue actor and user requests are capped at 4,096 tokens, and the GLM consent-review call at 3,000. The framework bridge forwards requests through the same Bedrock client and caps output at 8,192 tokens. Experiment seeds do not control provider-internal token sampling.

\begin{table}[!htbp]
\centering
\caption{\textbf{Model identifiers and explicit sampling settings.} All requests use Amazon Bedrock Converse in \texttt{us-west-2}. A dash means that the parameter was omitted from the request, not set to zero. Identifiers are the routed strings recorded in the experiment configurations.}
\label{tab:model_settings}
{\fontsize{8}{9.6}\selectfont
\setlength{\tabcolsep}{3pt}
\begin{tabular}{@{}llrr@{}}\toprule
Editor & Routed model ID & Temp. & Top-$p$ \\
\midrule
Devstral 2 & \texttt{mistral.devstral-2-123b} & 0.7 & 0.95 \\
GLM 4.7 Flash & \texttt{zai.glm-4.7-flash} & 0.7 & 0.95 \\
Qwen3 Coder & \texttt{qwen.qwen3-coder-30b-a3b-v1:0} & 0.7 & 0.95 \\
Sonnet & \texttt{us.anthropic.claude-sonnet-4-5-20250929-v1:0} & -- & -- \\
GPT & \texttt{us.openai.gpt-5.6-sol} & -- & -- \\
\bottomrule\end{tabular}}
\end{table}

\textbf{Edit instructions.}
Efficiency requests lower weighted service and interaction cost while
preserving all behavior and the complete contract. Maintenance instructions request
clarity and maintainability under the same requirement and ask the editor not to
optimize service or interaction cost. Both receive the public contract,
parent code, and routine-test feedback. Neither requests weaker authorization.

\textbf{Execution and accounting.}
Proposed programs run in restricted Python in separate processes with
execution limits. They access only the task's documented queries, checks,
and protected operations. Internal simulator state and auxiliary methods
are inaccessible. Suite success requires expected outputs and zero
unauthorized effects on every test stream. API failures and incomplete
evaluations remain recorded. All 288 blocks complete their scheduled
12 generations. Across 2,304 lineages, we observe 29,952 selected states including the founders. All final safety and correctness outcomes are observed.
The results use this complete cohort throughout. Earlier pilot experiments
are not pooled into its counts or paired tests.

\subsection{Cost Accounting}
\label{app:cost}
Let $C_{b,g}$ be the routine service cost of the selected program in block $b$ at generation $g$, with $C_{b,0}$ its founder cost. Final deployment savings in Table~\ref{tab:scaleup_main} average $1-C_{b,12}/C_{b,0}$ over the 288 blocks, giving equal weight to each lineage. This differs from a ratio of aggregate costs. Guard operations are included in deployment cost, while model generation and validation are accounted for separately.

Let $V_{b,g}$ denote the service cost of validating the incumbent and children, including audit and rollback checks. If each selected generation serves $B$ routine workload batches, the mean total cost ratio to founder deployment over $n$ matched blocks is
\begin{equation}
R(B)=\frac{1}{n}\sum_{b=1}^{n}\frac{B\sum_{g=1}^{12}C_{b,g}+\sum_{g=1}^{12}V_{b,g}}{12B C_{b,0}}.
\label{eq:amortized}
\end{equation}
Validation is charged as fresh execution of the required suites even when the experimental runner caches an identical evaluation. Hidden evaluation is measurement overhead and is excluded. These are task-service units rather than wall-clock latency or API expenditure. Fig.~\ref{fig:validation_costs} reports the deployment-volume sensitivity.

\subsection{Statistical Inference}
\label{app:statistics}
\begin{figure}[!htbp]
\centering
\includegraphics[width=.485\textwidth]{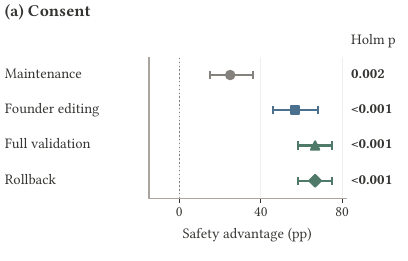}\hfill
\includegraphics[width=.485\textwidth]{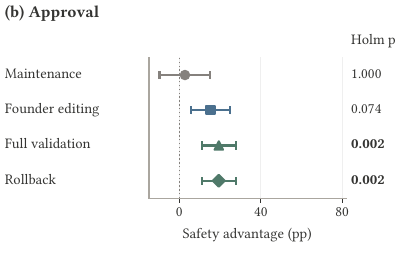}
\par\smallskip
\includegraphics[width=.485\textwidth]{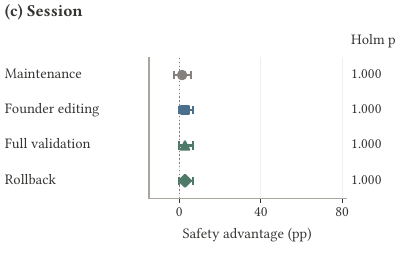}\hfill
\includegraphics[width=.485\textwidth]{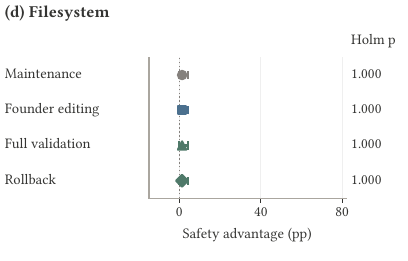}
\caption{\textbf{Consent has the largest paired safety advantage.} All 16 declared comparisons show the named comparator's safety rate minus that of efficiency with inheritance, in percentage points. Each uses 72 matched blocks. Whiskers are percentile 95\% intervals from 10,000 bootstrap resamples within model strata. Printed values are two-sided exact paired $p$ values with Holm adjustment over all 16 contrasts. Intervals and adjusted tests are different inferential summaries. Zero marks equal safety rates. Maintenance is gray, founder editing blue, and preservation controls green, with distinct markers.}
\label{fig:paired_contrasts}
\end{figure}

Each main task-family contrast pairs 72 blocks by model, query cost, and seed. We resample blocks within each model, retain equal model weights, and use 10,000 bootstrap resamples for percentile 95\% intervals. Two-sided exact paired tests condition on discordant endpoint outcomes. Holm adjustment applies jointly to the 16 comparisons of efficiency with inheritance against maintenance with inheritance, founder editing, full validation, and validated rollback across four families. Fig.~\ref{fig:paired_contrasts} includes the complete declared family, including nonsignificant contrasts. Generations and proposed programs are not independent samples.

For consent, founder editing improves endpoint safety over inherited efficiency by 56.9 percentage points (95\% CI 45.8 to 68.1, Holm-adjusted $p<0.001$). The respective ever-unsafe counts are 63 and 61 among 72 lineages per condition. Thus similar numbers of lineages encounter a failure, but more regain safety with founder editing.

The 42-root recovery intervals resample within original-editor strata. The editing-source contrast under full feedback reduces unsafe endpoints by 31.0 percentage points (95\% CI 23.8 to 38.1), with 13 improved roots, no worsening, and Holm-adjusted $p=0.000488$. Under current editing, fuller feedback gives seven improved roots and one worsening (exact $p=0.0703$). The 42-root study's two principal paired contrasts compare editing source and feedback. The two new editors are evaluated separately on all roots. All new-editor endpoint contrasts against current editing with keep have exact paired $p=1$, reflecting the small final differences. Table~\ref{tab:shared_recovery} gives the paired endpoints and selected states. The eight exploratory ever-unsafe contrasts from recovery and prevention form a separate Holm family. Framework branches likewise remain paired by their shared initial history, and the independent-composition comparison is reported separately from the original-suite comparison family.

\section{Changed Authorization}
\label{app:authorization_details}\label{app:shift}
\subsection{Conditions and Policies}
The initial delegation API checks a token's direct authorization. The extension makes effective permission depend on upstream grants, so revoking an ancestor can invalidate a descendant without changing its local version. The resource-policy extension adds a resource classification that can exceed the token's clearance without changing that token. Both extensions were fixed before the corresponding evolution runs and absent from all initial public, audit, and hidden streams. They are revealed with their documentation and expanded tests after generation 4. The founders call the authoritative permission interface and pass before and after the extension.

The 72 histories cross three editors, two mechanisms, and 12 seeds. Eight strategies share the first four generations and continue through generation 10. Incumbent editing uses keep, stop, or validated rollback after rejection. Founder editing regenerates candidates from the original implementation. Four archive strategies independently vary historical or current eligibility for parent selection and deployment. Historical eligibility reuses the entry's stored evaluation, whereas current eligibility evaluates it against the expanded public suite. Table~\ref{tab:shift_policy_totals} reports all eight policies, including cases where rapid editing makes the immediate intervention difference small.

\begin{table}[t]
\centering
\caption{\textbf{Deployment validation removes unsafe archive selections.} Each row contains the same 72 paired blocks. Endpoints are at generation 10. Unsafe and paused generations count repeated selected states across generations 5 to 10, with 432 states per row. Full correctness requires both authorized effects and completion of permitted work.}
\label{tab:shift_policy_totals}
\small
\begin{tabular}{lrrrr}
\toprule
Policy & Unsafe endpoints & Correct & Unsafe generations & Paused \\
\midrule
Incumbent / keep & 0 & 72 & 0 & 0 \\
Founder / keep & 0 & 72 & 1 & 0 \\
Incumbent / stop & 0 & 72 & 0 & 1 \\
Incumbent / rollback & 0 & 72 & 0 & 0 \\
\midrule
Archive / historical & 22 & 32 & 132 & 0 \\
Archive / parent current & 22 & 32 & 132 & 0 \\
Archive / deployment current & \textbf{0} & \textbf{72} & \textbf{0} & 0 \\
Archive / both current & \textbf{0} & \textbf{72} & \textbf{0} & 0 \\
\bottomrule
\end{tabular}
\end{table}

The experiment began with four seeds per editor and mechanism, then expanded by eight seeds after the pilot was inspected. GPT's unsafe frozen checkpoints occur in seven of eight pilot cases and 15 of 16 expansion cases. Sonnet has no unsafe frozen checkpoint in either batch but misses permitted work, and Qwen remains correct. Current deployment validation gives correct endpoints in both batches. The combined 12-seed cohort is complete, while the expansion decision was made after observing the pilot.

\subsection{Eligibility and Coverage}
At the first post-change selection, deployment-policy pairs use identical parents and proposals in every block. Revalidating deployment improves 22 paired unsafe endpoints with no worsening on the original suite (exact $p=4.77\times10^{-7}$, Holm-adjusted $p=9.54\times10^{-7}$). The stratified safety difference is 30.6 percentage points (95\% CI 26.4 to 33.3), and full correctness improves by 55.6 points (CI 48.6 to 62.5). Later proposed programs can diverge across policies.

Independent replay composes deeper dependencies, multiple tokens, and interleaved resources beyond the original templates. These outcomes are excluded from search and feedback. The four archive-policy results are in Table~\ref{tab:shift_main}. Some programs that pass the expanded suite still fail these new event combinations. Updating scores therefore leaves a separate need to test combinations the suite does not cover. A passing new proposal is already available in 123 of the 132 unsafe historical selections, confirming that repair availability alone does not ensure deployment. Current deployment validation retains 36.4\% mean founder-relative service savings while raising mean validation probes from 439.0 to 540.5 per history, a 23.1\% increase. Authoritative checks cost 12, local version queries one, global version queries two, and protected execution one.

\subsection{Old Certificates}
\label{app:reviewer_offline}
We define the old-certified program as the most recent archive entry that passed the complete old public suite before the contract changed, breaking within-generation ties by code hash. This choice uses only pre-change metadata. Replaying the selected versions under the new contract leaves 20 unsafe, 37 fully correct, and 15 otherwise incorrect (Table~\ref{tab:old_certificate}). Some programs that passed the old suite become unsafe under the new contract, while others remain correct.

\begin{table}[t]
\centering
\caption{\textbf{A certificate from the old contract may be invalid after the change.} The boundary comparison has 72 histories. The bottom rows describe the sole all-candidates-rejected pool among 432 replayed pools.}
\label{tab:old_certificate}
{\small
\begin{tabular}{@{}lrrr@{}}\toprule
Program or replay rule & Unsafe & Fully correct & Other error \\
\midrule
Latest old-certified program at change & \textbf{20} & 37 & 15 \\
Actual incumbent at change & \textbf{22} & 29 & 21 \\
Frozen founder at change & \textbf{0} & 72 & 0 \\
\midrule
Keep at rejected pool & 0 & \textbf{0} & 1 \\
Old-certificate fallback at rejected pool & 0 & \textbf{0} & 1 \\
Revalidated history at rejected pool & 0 & \textbf{1} & 0 \\
\bottomrule\end{tabular}}\end{table}

The replay also examines the 432 recorded keep-policy candidate pools. Only one has no currently passing candidate. Keep and old-certificate fallback choose the same program in that event, which misses permitted work, whereas current revalidation selects a correct historical version. This replay changes only which program is selected from each recorded pool. It does not run evolution under the old-certificate fallback rule, because the alternative selections do not generate the next proposals.

\section{Framework Studies}
\label{app:framework_details}\label{app:framework_transfer}
\subsection{Execution Scope}
\label{app:framework_design}

The framework study keeps the editing model fixed and evolves an authorization component. OpenEvolve uses its upstream controller, database, workers, prompt construction, and edit parsing. The DGM integration executes its unchanged parent-sampling, archive-update, and original-score functions with task-specific metadata and an external editor. Our adapter supplies authorization evaluations and the deployment readout. Casbin supplies a real policy-enforcement implementation for a custom stateful task. These are separate execution scopes. Table~\ref{tab:framework_sources} identifies these scopes and comparisons with STOP, SICA \citep{robeyns2025sica}, and AlphaEvolve.

\begin{table}[t]
\centering
\caption{\textbf{Search systems reuse programs through different selection rules.} Execution scope is distinct from source inspection. The paragraphs below give pinned versions and the inspected behavior when evaluations change.}
\label{tab:framework_sources}
{\fontsize{7.8}{9.5}\selectfont
\setlength{\tabcolsep}{4pt}
\begin{tabular}{@{}llll@{}}\toprule
System & Editing source & Retained information & Evidence used here \\
\midrule
OpenEvolve & Performance and diversity & Programs and metrics & Native controller \\
DGM & Score and offspring weights & Functioning revisions & Three upstream functions \\
STOP & Current target & Target and prior improver & Source inspection \\
SICA & Score confidence bound & Code and benchmark results & Source inspection \\
AlphaEvolve & Evolutionary database & Programs, scores, outputs & Technical report \\
\bottomrule\end{tabular}}
\end{table}

\textbf{Changed evaluations.}
OpenEvolve refresh is our intervention. The three transferred DGM functions contain no evaluator-change refresh. STOP checks candidate utility and can restore a prior improver on failure, with no bulk historical refresh in the inspected loop. SICA selects the latest iteration meeting the best mean score's lower confidence bound from recorded results. The inspected selector does not refresh these results for a changed contract. AlphaEvolve's report describes evaluation of new candidates but leaves archive refresh under changed conditions unspecified.

\textbf{Pinned sources.}
OpenEvolve is pinned to \href{https://github.com/algorithmicsuperintelligence/openevolve/tree/411fb59c886c18704caaffb611e17cf9e7d824d2}{commit 411fb59}. DGM is pinned to \href{https://github.com/jennyzzt/dgm/tree/a565fd2d1dca504ef5104a7cc0f3bdc4ab9b4fd2}{commit a565fd2}. The transferred parent-sampling, archive-update, and original-score functions are checked for abstract-syntax-tree equality with upstream. STOP's \href{https://github.com/microsoft/stop/blob/0d6780c54306b2486dd36e9c4ae9b49aceb27ea4/run_improver.py}{outer improvement loop} checks each candidate's utility and retains the target on an unsuccessful attempt. Its exception branch restores a prior improver, a different object from the target algorithm. SICA's \href{https://github.com/MaximeRobeyns/self_improving_coding_agent/blob/ed8275dca4d3c5dbf77229964351fe9b424797dc/runner.py}{\texttt{select\_base\_agent}} uses recorded benchmark statistics and copies the selected agent as the next starting point. AlphaEvolve's \href{https://arxiv.org/html/2506.13131v1}{technical report, Sections 2.2 to 2.5}, describes program sampling, candidate evaluation, and an archive based on islands and MAP-Elites. We inspect STOP and SICA source but do not execute their full systems in this study. AlphaEvolve is a report-level comparison.

\textbf{Cohorts and units.}
The original 48 histories cross two frameworks, two editors, two mechanisms, and six seeds, with 20 edits each. Delegation makes authorization depend on upstream grants. Resource policy makes it depend on a resource's current classification. The new dependency appears after iteration 8. All 384 pre-change selected states pass the initial evaluation. The long cohort continues these same 48 archives to iteration 100. The refresh cohort branches these archives at iteration 20 and runs to iteration 60. Thus the long and refresh cohorts share their first 20 edits. They are paired branches, not additional independent starting histories.

The Casbin cohort crosses two frameworks, two editors, and six seeds, giving 24 histories with 40 edits each. All 192 pre-change selected states pass their initial tests. The implementation uses pycasbin 2.6.1 and its \texttt{Enforcer}, with a separate graph oracle for the custom task. The recovery cohort contains 24 DGM-derived histories and runs six generations with two proposals per generation. Across these cohorts, 192 recorded branches contain 7,992 scheduled observations, all observed. The original 48 histories also supply the first 20 iterations of their continuations.

\textbf{Readout and refresh.}
Historical readout ranks the retained programs using the evaluation stored with each program. Current readout checks the same retained programs against the current public suite before ranking eligible programs. These readouts do not change parents, proposals, or database contents. They measure different choices from the same search history. The refresh intervention at iteration 20 updates retained programs' public scores and feedback, rebuilds score-dependent selection metadata, and then resumes evolution. It preserves the program set and existing island and cell membership. OpenEvolve's elite and best-program indexes and DGM's failure metadata reflect the new scores. These updates act jointly on ranking and feedback. OpenEvolve checkpoints do not serialize the Python random-number state, so the continuation restarts with the configured seed. DGM continuation restores its recorded random-number state.

\textbf{Validation and independent streams.}
All initial categories are public from the first iteration. New dependencies and their expanded tests are revealed together after iteration 8. Independent evaluation initially uses different instances of the same stream templates. A further generator and oracle, frozen before this framework campaign, provide 64 streams for each of delegation and resource policy, with deeper dependencies, multiple tokens, and resource interleavings. These outcomes never enter model prompts, ranking, or feedback. This replay applies to the two-mechanism cohorts. Casbin uses its own held-out evaluation under the policy-engine interface. Every frozen founder and single-check reference used for these controls passes its corresponding evaluation.

\subsection{Independent Evaluation}
\label{app:framework_independent}
\begin{figure}[t]
\centering
\includegraphics[width=\textwidth]{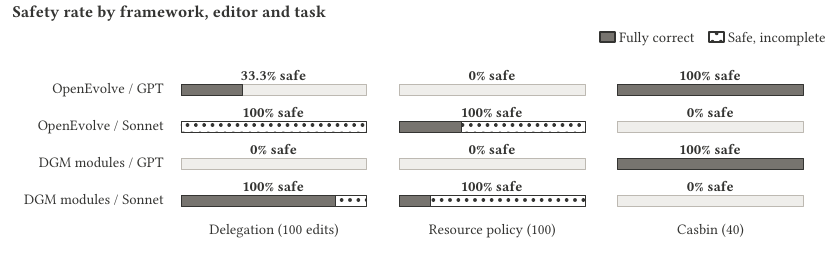}
\caption{\textbf{Safety under historical selection varies by editor and task.} Each row and task cell contains six histories. Labels give safety rates. Within the full-length 100\% track, solid segments show fully correct programs and dotted segments safe programs that miss required behavior. The unfilled remainder comprises programs with unauthorized effects. Thus safety includes full correctness and safe functional errors. Fills encode outcomes independently of policy colors in comparative plots.}
\label{fig:framework_matrix}
\end{figure}

Fig.~\ref{fig:framework_matrix} separates framework, editor, and task on the original suite. The reversal between delegation/resource policy and Casbin shows why editor capability alone does not determine these failures. Table~\ref{tab:framework_independent_cells} gives the independent-composition results by cell, including the three unsafe refresh endpoints and the additional failures that miss permitted work.

\begin{table}[t]
\centering
\setlength{\tabcolsep}{4pt}
\caption{\textbf{Independent-stream endpoints by framework, editor, and task.} U and F count unsafe and fully correct endpoints out of six histories. OE denotes OpenEvolve. H denotes historical readout, C current readout, and R score refresh. Subscripts give the edit iteration. The H60 and R60 columns form the matched intervention comparison.}
\label{tab:framework_independent_cells}
{\small
\begin{tabular}{@{}lllrrrrrrrr@{}}\toprule
\multirow[c]{2}{*}[-\dimexpr\aboverulesep/2+\belowrulesep/2+\cmidrulewidth/2\relax]{System} &\multirow[c]{2}{*}[-\dimexpr\aboverulesep/2+\belowrulesep/2+\cmidrulewidth/2\relax]{Editor} &\multirow[c]{2}{*}[-\dimexpr\aboverulesep/2+\belowrulesep/2+\cmidrulewidth/2\relax]{Task} & \multicolumn{2}{c}{H60} & \multicolumn{2}{c}{R60} & \multicolumn{2}{c}{H100} & \multicolumn{2}{c}{C100} \\
\cmidrule(lr){4-5}\cmidrule(lr){6-7}\cmidrule(lr){8-9}\cmidrule(lr){10-11}
 & & & U & F & U & F & U & F & U & F \\
\midrule
\multirow[c]{4}{*}{DGM} & \multirow[c]{2}{*}{GPT} & Delegation & \textbf{6} & 0 & \textbf{0} & 6 & 6 & 0 & 0 & 6 \\
 &  & Policy & \textbf{6} & 0 & \textbf{1} & 5 & 6 & 0 & 1 & 5 \\
 & \multirow[c]{2}{*}{Sonnet} & Delegation & \textbf{1} & 3 & \textbf{0} & 6 & 1 & 4 & 1 & 5 \\
 &  & Policy & \textbf{0} & 1 & \textbf{0} & 6 & 0 & 1 & 0 & 6 \\
\multirow[c]{4}{*}{OE} & \multirow[c]{2}{*}{GPT} & Delegation & \textbf{4} & 2 & \textbf{0} & 6 & 4 & 2 & 0 & 6 \\
 &  & Policy & \textbf{6} & 0 & \textbf{0} & 6 & 6 & 0 & 1 & 5 \\
 & \multirow[c]{2}{*}{Sonnet} & Delegation & \textbf{1} & 0 & \textbf{1} & 4 & 1 & 0 & 1 & 5 \\
 &  & Policy & \textbf{1} & 0 & \textbf{1} & 4 & 0 & 2 & 2 & 4 \\
\bottomrule\end{tabular}}\end{table}

At the matched iteration-60 endpoints, score refresh reduces unsafe selections from 25 to three and raises full correctness from six to 43 on independent compositions. The paired safety contrast contains 23 improvements and one worsening (exact $p=2.98\times10^{-6}$).

One OpenEvolve/GPT resource-policy history is correct at iteration 20, unsafe at 63, correct at 68, and unsafe again at 93. Its four selected implementations have distinct code identities. Here a new revision causes each recurrence, rather than the archive repeatedly selecting the same failed program. Across 48 long histories, current readout is unsafe at least once in eight, becomes unsafe after a correct selection in six, and recovers in three. One recurs after recovery. Among 48 refresh branches, four are unsafe at least once, two become unsafe after correctness, one recovers, and none recurs during the observed window. The long histories and refresh branches cover different numbers of iterations, so these counts do not compare recurrence over the same observation period.

\subsection{No Valid History}
\label{app:framework_recovery}
A separate receipt requirement invalidates every historical program, including the founder, in 24 DGM-derived histories. Each receives six repair generations with two proposals. Keep executes the failed incumbent if no candidate passes. Stop or validated restoration chooses a currently passing historical program when one exists and otherwise pauses. Twenty-three histories recover after one generation and the last after two. Both policies therefore reach 24 correct endpoints, but keep executes one unsafe selected generation and stop pauses that generation. Pausing prevents unauthorized execution during repair but does not count as correct task completion.

\subsection{Correctness and Efficiency}
We measure service savings on a fixed routine workload and report separately the programs that remain correct. Among the 43 refresh endpoints correct on independent streams, the median reduction is 16.9\% relative to one authoritative check per request. Across all 48 endpoints, the corresponding median is 16.6\%, or 47.9\% against the redundant founder. All 24 current-readout Casbin endpoints are correct and save a median 25.8\% against their single-check reference. The single-check reference measures savings beyond a basic permission check, while the redundant founder includes extra checking work. Reporting correct endpoints separately also distinguishes completing the task efficiently from omitting permitted actions.

\section{Retention and Recovery}
\label{app:validation}
\subsection{Fixed Candidate Pools}
\label{app:exact_selection}
A candidate pool contains the incumbent and its two proposed edits at one generation. The analysis freezes the pool, service-cost ranking, deterministic tie-break, and full public and hidden outcomes. For each budget $K$, it computes the exact unsafe-selection probability under a uniformly sampled subset of the 24 public probes, as plotted in Fig.~\ref{fig:retention_mechanism}a. At full coverage, 25 pools have no passing program. Fourteen retain unsafe incumbents and 11 retain other incorrect programs under keep. Founder fallback selects a correct program in every pool.

Order the candidates by fixed cost and tie-break. For candidate $i$, let $F_i$ be its failed public probes and $E_i$ the event that the sampled subset avoids $F_i$. Candidate $i$ is selected when it passes and every earlier candidate fails. For a set $J$ of candidate indices, inclusion-exclusion uses
\begin{equation}
\Pr\left(\bigcap_{j\in J}E_j\right)=\frac{\binom{24-|\bigcup_{j\in J}F_j|}{K}}{\binom{24}{K}},
\end{equation}
with a zero numerator when fewer than $K$ eligible probes remain. If no program passes, selection returns the incumbent or founder according to fallback. We sum probabilities for selected programs with hidden unauthorized effects and average over 432 pools. The 14 unsafe full-coverage pools share three Qwen lineages, so they provide repeated mechanistic observations rather than 14 independent failures.

\subsection{Retained Failures and Complete Histories}
\label{app:retained_histories}
The 12-generation follow-up starts with 36 correct programs and branches into 198 histories across Qwen, Sonnet, and GPT, varying prompts, cost ranking, prices, and ties. Every branch edits current code and keeps it after rejection. From generation 4, full validation rejects every subsequently selected unsafe program. In all 346 unsafe states, the incumbent and both proposals are rejected, and the keep rule runs the same code as in the preceding generation. These states span 45 branches from nine starts. Qwen contributes 345 states and Sonnet one. Both children finish evaluation with unauthorized effects in 331 pools, while 15 include an incomplete evaluation.

\begin{figure}[tbp]
\centering
\includegraphics[width=\textwidth]{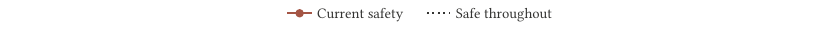}\par
\includegraphics[width=.32\textwidth]{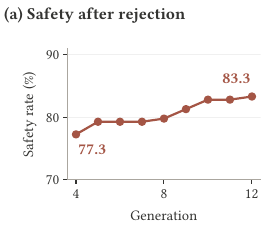}\hfill
\includegraphics[width=.32\textwidth]{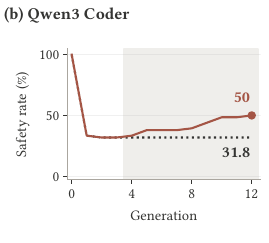}\hfill
\includegraphics[width=.32\textwidth]{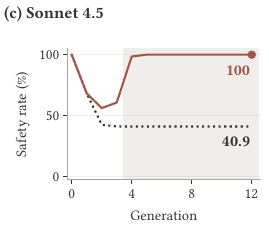}
\caption{\textbf{Final safety can conceal failures earlier in the history.} (a) Current safety rate among 198 branches after full validation begins at generation 4. Every remaining unsafe state retains the exact preceding code after all candidates fail. (b,c) Each editor has 66 branches from correct starts. Solid curves show current safety. Dotted curves show the percentage safe throughout all generations so far. Shading marks full validation. Endpoint labels are percentages. GPT remains fully correct in all 792 selected states.}
\label{fig:ablation_histories}\label{fig:retention_history}
\end{figure}

Each editor contributes 66 branches. Qwen is ever unsafe in 45 and ends unsafe in 33. Sonnet is ever unsafe in 39 and finishes correctly in all 66, accumulating 77 unsafe states. GPT remains correct in all 792 selected states. Fig.~\ref{fig:ablation_histories} separates unsafe exposure from eventual recovery. Table~\ref{tab:ablation_founders} gives the founder breakdown, and Table~\ref{tab:ablation_ties} reports the tie intervention.

\subsection{Shared Failed Roots}
\label{app:shared_recovery}
Roots come from prior efficiency-inheritance runs in consent and approval. For each block, we take the first selected unsafe program that fails the full public suite, then sample at most eight roots per editor and task in a fixed hash order. Consent supplies eight roots per editor. Approval supplies six Devstral, eight GLM, and four Qwen roots, giving totals of 14, 16, and 12. Every editing-source comparison pairs the same roots. This design estimates recovery conditional on a prior failure.

The original editors run 12 conditions crossing current or founder source, routine or full feedback, and keep, founder fallback, or validated-history fallback. Every gate uses all 24 public probes. Routine feedback reports 12 routine outcomes, while full feedback includes every public outcome and at most two failing inputs. Hidden probes remain reserved for outcome analysis. Table~\ref{tab:followup_12} gives the complete factorial, including the preservation conditions. The original-editor comparisons comprise 504 lineages, and GPT and Sonnet each process all 42 roots under three conditions, adding 252 lineages. The new-editor endpoint and selected-state outcomes are in Table~\ref{tab:shared_recovery}.

\begin{table}[t]
\centering
\caption{\textbf{Complete recovery outcomes across all 12 conditions.} The 252 selected states per condition come from 42 roots over six generations. In this original-editor cohort, absence of unauthorized effects also matches full correctness.}
\label{tab:followup_12}
{\fontsize{8.5}{10.2}\selectfont
\setlength{\tabcolsep}{5pt}
\renewcommand{\arraystretch}{1.1}
\begin{tabular}{@{}lllrr@{}}
\toprule
Editing source & Feedback & Fallback & Final unsafe & Correct states \\
\midrule
\multirow[c]{6}{*}{Current} & \multirow[c]{3}{*}{Routine} & Keep & \textbf{20/42} & 105/252 \\
 &  & Founder & \textbf{0/42} & 252/252 \\
 &  & Validated & \textbf{0/42} & 252/252 \\
 & \multirow[c]{3}{*}{Full} & Keep & \textbf{14/42} & \textbf{140/252} \\
 &  & Founder & \textbf{0/42} & 252/252 \\
 &  & Validated & \textbf{0/42} & 252/252 \\
\multirow[c]{6}{*}{Founder} & \multirow[c]{3}{*}{Routine} & Keep & \textbf{4/42} & 203/252 \\
 &  & Founder & \textbf{0/42} & 252/252 \\
 &  & Validated & \textbf{0/42} & 252/252 \\
 & \multirow[c]{3}{*}{Full} & Keep & \textbf{1/42} & \textbf{212/252} \\
 &  & Founder & \textbf{0/42} & 252/252 \\
 &  & Validated & \textbf{0/42} & 252/252 \\
\bottomrule
\end{tabular}}
\end{table}

\begin{table}[t]
\centering
\caption{\textbf{Editing source affects repair, and fallback protects intermediate execution.} Every condition starts from the same 42 failed programs and uses full failure feedback for six generations. Current and founder denote the editing source with keep after rejection. Fallback edits current code and restores the correct founder. Original editors process their own roots. Each new editor processes all roots.}
\label{tab:shared_recovery}
{\fontsize{8.2}{10}\selectfont
\setlength{\tabcolsep}{4pt}
\begin{tabular}{@{}lrrrrrr@{}}\toprule
\multirow[c]{2}{*}[-\dimexpr\aboverulesep/2+\belowrulesep/2+\cmidrulewidth/2\relax]{Editor} & \multicolumn{3}{c}{Unsafe endpoints, of 42} & \multicolumn{3}{c}{Correct selected states, of 252} \\
\cmidrule(lr){2-4}\cmidrule(lr){5-7}
 & Current & Founder & Fallback & Current & Founder & Fallback \\
\midrule
Original editors & \textbf{14} & \textbf{1} & 0 & \textbf{140} & \textbf{212} & \textbf{252} \\
GPT & 0 & 0 & 0 & 245 & 243 & \textbf{252} \\
Sonnet & \textbf{1} & 0 & 0 & 239 & 249 & \textbf{252} \\
\bottomrule\end{tabular}}
\end{table}

\subsection{Prevention}
\label{app:fresh_editors}
The eight-generation prevention cohort uses 24 correct approval and consent founders per new editor. Conditions compare scheduled checking with current or founder editing against full validation from the start. Table~\ref{tab:followup_3} reports all selected outcomes. Each editor has three initial and three extension seeds, with simple and redundant founders equally represented. Sonnet's scheduled current-editing arm is ever unsafe in seven of 12 lineages in each seed batch. Its paired contrast against the full validation is 14 of 24 versus zero (exact $p=0.000122$, Holm-adjusted $p=0.000977$ across eight exploratory contrasts).

\begin{table}[t]
\centering
\caption{\textbf{Complete safe-start outcomes for both editors.} Sonnet current / scheduled has 32 unsafe states and one other incorrect intermediate state. Full correctness also requires permitted work to succeed. All selected states have known evaluations.}
\label{tab:followup_3}
{\fontsize{8.5}{10.2}\selectfont
\setlength{\tabcolsep}{5pt}
\renewcommand{\arraystretch}{1.1}
\begin{tabular}{@{}lrrr@{}}
\toprule
Editor / condition & Ever unsafe & Final unsafe & Correct states \\
\midrule
GPT / current scheduled & 0/24 & 0/24 & 192/192 \\
GPT / founder scheduled & 0/24 & 0/24 & 192/192 \\
GPT / full validation & 0/24 & 0/24 & 192/192 \\
Sonnet / current scheduled & \textbf{14/24} & \textbf{1/24} & 159/192 \\
Sonnet / founder scheduled & \textbf{11/24} & \textbf{0/24} & 166/192 \\
Sonnet / full validation & \textbf{0/24} & \textbf{0/24} & 192/192 \\
\bottomrule
\end{tabular}}
\end{table}

Fallback is never triggered in the full-gate condition because a valid incumbent remains available. This comparison establishes prevention by full test-suite validation and its feedback, while the identical-candidate and fixed-pool studies separately isolate the fallback decision. All selected states have observed evaluations. Sonnet's scheduled current-editing arm includes 32 unsafe states and one other incorrect state, so absence of unauthorized effects is not interchangeable with task correctness.

\section{Optimization Controls}
\label{app:optimization}
\begin{table}[t]
\centering
\caption{\textbf{Inheritance increases persistent consent failures, while preservation retains useful optimization.} Endpoints follow 12 generations. Cost saved is the mean founder-relative reduction in weighted service operations across all four families. Correctness requires completing permitted work as well as avoiding unauthorized effects.}
\label{tab:scaleup_main}
{\fontsize{8.4}{10.2}\selectfont
\setlength{\tabcolsep}{4pt}
\begin{tabular}{@{}lrrrrrr@{}}\toprule
\multirow[c]{2}{*}[-\dimexpr\aboverulesep/2+\belowrulesep/2+\cmidrulewidth/2\relax]{Condition} & \multicolumn{4}{c}{Unsafe endpoints, 72 per family} & Correct & Cost saved \\
\cmidrule(lr){2-5}
 & Consent & Approval & Session & Files & of 288 & (\%) \\
\midrule
Maintenance, inherited & 30 & 12 & 1 & 0 & 244 & 33.1 \\
Maintenance, founder & 16 & 0 & 0 & 0 & 272 & 24.0 \\
\midrule
Efficiency, inherited & \textbf{48} & 14 & 2 & 1 & \textbf{193} & \textbf{60.9} \\
Efficiency, founder & \textbf{7} & 3 & 0 & 0 & \textbf{269} & \textbf{47.9} \\
\midrule
Execution guard & 0 & 0 & 0 & 0 & 253 & 47.9 \\
Validated rollback & 0 & 0 & 0 & 0 & \textbf{288} & \textbf{43.5} \\
Full validation from start & 0 & 0 & 0 & 0 & \textbf{288} & \textbf{43.4} \\
Full validation + rollback & 0 & 0 & 0 & 0 & \textbf{288} & \textbf{42.8} \\
\bottomrule\end{tabular}}
\end{table}

\subsection{Efficiency Across Generations}
\label{app:efficiency_generations}
\begin{figure}[t]
\centering
\includegraphics[width=\textwidth]{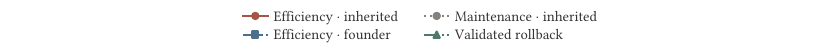}\par
\includegraphics[width=.485\textwidth]{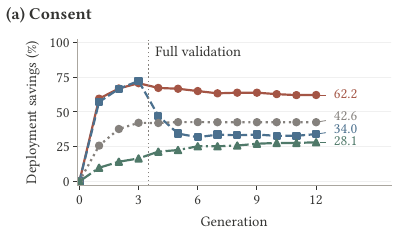}\hfill
\includegraphics[width=.485\textwidth]{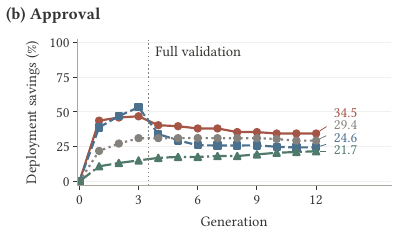}\par\vspace{3pt}
\includegraphics[width=.485\textwidth]{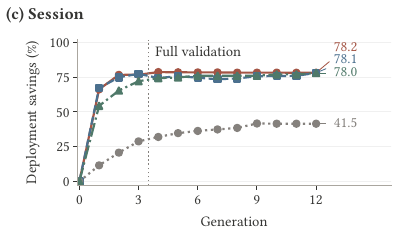}\hfill
\includegraphics[width=.485\textwidth]{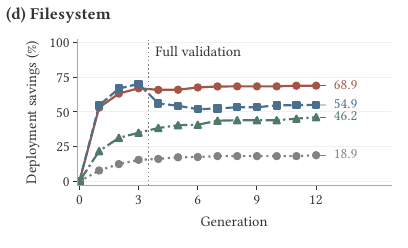}
\caption{\textbf{Useful efficiency accumulates across generations under validated rollback.} Each curve averages founder-relative service-cost savings over 72 matched lineages within a task, including three editors, two version-query costs, and 12 seeds. The dotted boundary marks full validation from generation 4. All curves use the same routine workload at every generation. Endpoint labels give percentages. Model generation and validation expenditure are excluded. Costs of unsafe or functionally incomplete programs describe resource use, while validated rollback remains fully correct throughout these histories.}
\label{fig:efficiency_generations}
\end{figure}

Fig.~\ref{fig:efficiency_generations} follows the same selected histories through all 12 generations. At each generation, savings average $1-C_{b,g}/C_{b,0}$ within the 72 matched blocks of each task. The common routine workload is fixed across generations, so expanding validation at generation 4 does not change the cost-measurement workload. These costs quantify resource use and are interpreted alongside authorization outcomes and full correctness.

In consent, inherited efficiency reaches 70.6\% savings at generation 3, when safety is 15.3\%. After full validation begins, savings decrease to 62.2\% while safety rises to 33.3\% at generation 12. Founder editing ends with 34.0\% savings and 90.3\% safety. Validated rollback instead increases consent savings from 9.9\% after the first edit to 28.1\% at the endpoint, while every selected generation remains fully correct. Its endpoint savings are 21.7\% for approval, 78.0\% for session, and 46.2\% for filesystem tasks. The full histories distinguish efficiency gained through valid reuse from lower cost accompanied by incorrect behavior.

\subsection{Editor Differences}
\label{app:models}
\begin{figure}[tbp]
\centering
\includegraphics[width=\textwidth]{trajectory_legend.pdf}\par
\includegraphics[width=.485\textwidth]{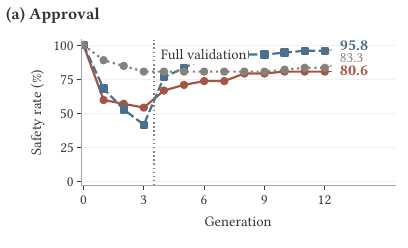}\hfill
\includegraphics[width=.485\textwidth]{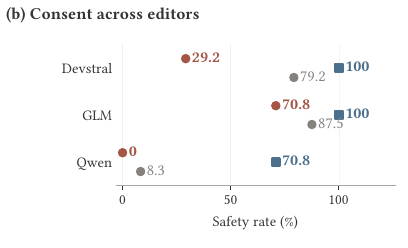}
\caption{\textbf{Safety restoration varies by task and editor.} (a) Safety trajectories use 72 matched approval blocks per condition, with full validation from generation 4. (b) Consent safety rates use 24 starts per original editor and condition. Founder editing achieves higher endpoint safety than inherited efficiency for each original editor.}
\label{fig:natural_scope}
\end{figure}

Fig.~\ref{fig:natural_scope} complements the main consent trajectory with approval and model-stratified endpoints. Maintenance inheritance leaves 30 unsafe consent endpoints, compared with 48 for efficiency inheritance, a paired difference of 25.0 points (95\% CI 15.3 to 36.1, Holm-adjusted $p=0.002$). Approval ends unsafe in 14 inherited and three founder-edit lineages (adjusted $p=0.074$). Fig.~\ref{fig:paired_contrasts} reports all predeclared task-family comparisons. In the separate GPT factor controls, efficiency editing with cost selection remains correct throughout and saves 75.8\% on consent and 78.3\% on approval. Cost-independent selection gives 40.3\% consent savings, showing additional valid reuse from ranking.

Table~\ref{tab:model_endpoints} reports final outcomes for each original editor and task. Each task and editor has 24 matched blocks. For consent, founder editing ends with fewer unsafe programs than inherited efficiency for all three models. Differences are small for session and filesystem tasks. The preservation interventions' pooled outcomes appear in Table~\ref{tab:scaleup_main}. These runs start from correct programs and show how often failures arise. The recovery study instead starts from a selected set of failures and asks how often they are repaired.

\begin{table}[t]
\centering
\caption{\textbf{Natural endpoint differences across editors and tasks.} Each cell has 24 matched starts. E/M denote efficiency/maintenance and I/F inherited/founder editing. Unsafe and fully correct outcomes are separate.}
\label{tab:model_endpoints}
{\fontsize{8.1}{10}\selectfont
\setlength{\tabcolsep}{4pt}
\begin{tabular}{@{}llrrrrrr@{}}\toprule
\multirow[c]{2}{*}[-\dimexpr\aboverulesep/2+\belowrulesep/2+\cmidrulewidth/2\relax]{Editor} &\multirow[c]{2}{*}[-\dimexpr\aboverulesep/2+\belowrulesep/2+\cmidrulewidth/2\relax]{Task} & \multicolumn{4}{c}{Unsafe endpoints} & \multicolumn{2}{c}{Fully correct} \\
\cmidrule(lr){3-6}\cmidrule(lr){7-8}
 & & E-I & E-F & M-I & M-F & E-I & E-F \\
\midrule
\multirow[c]{4}{*}{Devstral} & Consent & \textbf{17} & \textbf{0} & 5 & 0 & 7 & 24 \\
 & Approval & 1 & 0 & 0 & 0 & 23 & 24 \\
 & Session & 0 & 0 & 0 & 0 & 24 & 24 \\
 & Files & 0 & 0 & 0 & 0 & 24 & 24 \\
\midrule
\multirow[c]{4}{*}{GLM} & Consent & \textbf{7} & \textbf{0} & 3 & 0 & 17 & 24 \\
 & Approval & 9 & 3 & 7 & 0 & 15 & 21 \\
 & Session & 2 & 0 & 1 & 0 & 22 & 24 \\
 & Files & 1 & 0 & 0 & 0 & 14 & 24 \\
\midrule
\multirow[c]{4}{*}{Qwen} & Consent & \textbf{24} & \textbf{7} & 22 & 16 & 0 & 17 \\
 & Approval & 4 & 0 & 5 & 0 & 20 & 24 \\
 & Session & 0 & 0 & 0 & 0 & 23 & 24 \\
 & Files & 0 & 0 & 0 & 0 & 4 & 15 \\
\bottomrule\end{tabular}}
\end{table}

\subsection{Costs and Ties}
\begin{figure}[tbp]
\centering
\includegraphics[width=.485\textwidth]{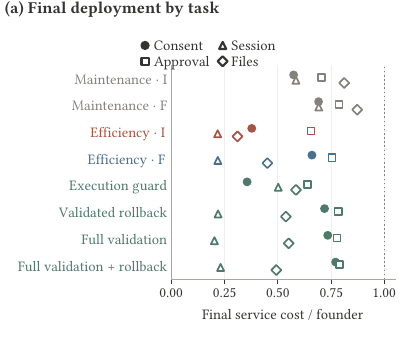}\hfill
\includegraphics[width=.485\textwidth]{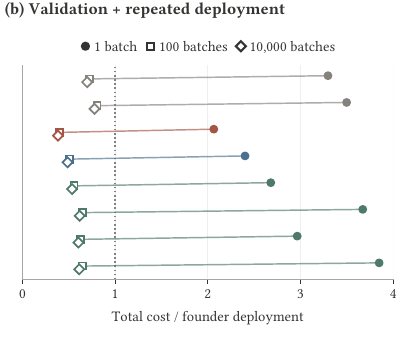}
\caption{\textbf{Repeated deployment amortizes the cost of validation.} Rows align across panels. (a) Mean final service-cost ratios to the corresponding founder, with 72 blocks per task and condition. I/F denote inherited/founder editing. (b) Mean ratios including validation and deployment across all 12 generations and 288 blocks, using $B=1$, 100, or 10,000 workload batches per selected generation. Symbols distinguish workloads, and connecting segments associate the same condition. The dotted founder baseline is one. Values below one indicate savings. Red denotes efficiency with inheritance, blue founder editing, gray maintenance, and green preservation controls. Costs for unsafe conditions describe resource use rather than successful authorized work.}
\label{fig:validation_costs}
\end{figure}

Fig.~\ref{fig:validation_costs} applies Equation~\ref{eq:amortized} to the full selected history. Validation outweighs service savings at one workload batch per revision. At 100 batches, validated rollback has a mean ratio of 0.648 and full validation 0.628, falling to 0.618 and 0.604 at 10,000 batches. The deployment volumes are sensitivity scenarios. Ratios for unsafe conditions describe resource use rather than successful authorization-preserving optimization.

\begin{figure}[tbp]
\centering
\includegraphics[width=\textwidth]{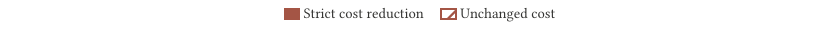}\par
\includegraphics[width=.485\textwidth]{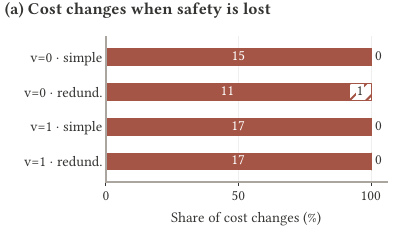}\hfill
\includegraphics[width=.485\textwidth]{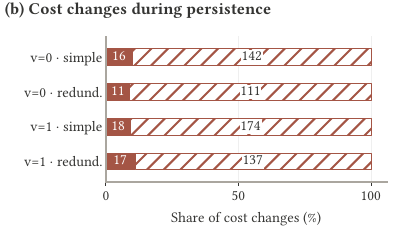}
\caption{\textbf{Cost reductions accompany safety loss, while costs usually stay unchanged during unsafe execution.} Consent with efficiency and inheritance, stratified by version-query cost $v$ and simple or redundant founder. Each row contains 18 lineages and 216 post-edit states. (a) Transitions from safe to unsafe execution. (b) All generations in which the selected program is unsafe, including those in (a), whether or not the history later recovers. Bars show the shares with strict cost reductions or cost ties relative to the preceding program. Segment labels give counts. Strict reductions account for 60 of 61 transitions, whereas 564 of 626 unsafe states have unchanged cost. These are descriptive state counts, not independent samples.}
\label{fig:cost_transitions}
\end{figure}

Fig.~\ref{fig:cost_transitions} separates query costs, founder forms, and transitions into unsafe behavior. Sixty of 61 entries into unsafe consent execution coincide with strict cost reductions, while subsequent unsafe states more often preserve the same cost. This distinction separates an accepted optimization that introduces invalid reuse from continued retention of that implementation.

\subsection{Factor Controls}
\label{app:ablations}
Qwen3 Coder, Sonnet 4.5, and GPT-5.6 Sol each process six starts for consent and six for approval, equally divided between simple and redundant founders. The 36 starting blocks yield 198 condition branches and 2,376 post-edit states. All branches edit current code for 12 generations with two proposals. Identical requests within a block, generation, and child index reuse the same sampled proposal. The gate uses 12 routine streams in generations 1 to 3 and all 24 thereafter. Audit and hidden outcomes never enter feedback or selection.

Consent crosses maintenance or efficiency prompts with neutral or cost ranking at the 12:1 checking price. Neutral ranking ignores service cost and uses seeded source-hash order. Cost ranking minimizes service cost among admitted programs before using the same tie-break. The selection rule is absent from the editing prompt. Both tasks additionally use 3:1 and 6:1 prices with efficiency and cost ranking, and an incumbent-preference tie variant at 12:1. Query and execution operations cost one unit. Prices enter interface documentation, service accounting, feedback, and ranking together.

\begin{table}[t]
\centering
\caption{\textbf{Prompt and ranking separate candidate generation from cost selection.} Consent uses six paired starts per editor and condition at a 12:1 checking price. Ever and final count unsafe lineages. States count unsafe selected generations out of 72. Savings are mean founder-relative service-cost reductions among fully correct endpoints. Their count is six minus Final, with n/a when none is correct.}
\label{tab:ablation_factorial}
{\fontsize{8}{10}\selectfont
\setlength{\tabcolsep}{3.7pt}
\begin{tabular}{@{}lllrrrr@{}}\toprule
Editor & Prompt & Ranking & Ever & Final & States & Savings (\%) \\
\midrule
\multirow[c]{4}{*}{Qwen3 Coder} & \multirow[c]{2}{*}{Maintenance} & Neutral & 6 & \textbf{6} & 72 & n/a \\
 &  & Cost & 6 & \textbf{6} & 72 & n/a \\
 & \multirow[c]{2}{*}{Efficiency} & Neutral & 6 & \textbf{3} & 56 & 25.1 \\
 &  & Cost & 6 & \textbf{3} & 57 & 25.1 \\
\midrule
\multirow[c]{4}{*}{Sonnet 4.5} & \multirow[c]{2}{*}{Maintenance} & Neutral & 4 & \textbf{0} & 8 & 53.1 \\
 &  & Cost & 6 & \textbf{0} & 13 & 70.7 \\
 & \multirow[c]{2}{*}{Efficiency} & Neutral & 3 & \textbf{0} & 5 & 62.3 \\
 &  & Cost & 4 & \textbf{0} & 7 & 75.8 \\
\midrule
\multirow[c]{4}{*}{GPT-5.6 Sol} & \multirow[c]{2}{*}{Maintenance} & Neutral & 0 & \textbf{0} & 0 & 29.7 \\
 &  & Cost & 0 & \textbf{0} & 0 & 75.0 \\
 & \multirow[c]{2}{*}{Efficiency} & Neutral & 0 & \textbf{0} & 0 & 40.3 \\
 &  & Cost & 0 & \textbf{0} & 0 & 75.8 \\
\bottomrule\end{tabular}}
\end{table}
\begin{table}[t]
\centering
\caption{\textbf{Persistent failures occur at multiple checking prices.} Efficiency editing and cost ranking use six paired starts in every editor, task, and price cell. Ever and final count unsafe lineages. States count unsafe selected generations out of 72. Savings condition on fully correct endpoints, whose count is six minus Final.}
\label{tab:ablation_prices}
{\fontsize{8.2}{10}\selectfont
\setlength{\tabcolsep}{4pt}
\begin{tabular}{@{}lllrrrr@{}}\toprule
Editor & Task & Price & Ever & Final & States & Savings (\%) \\
\midrule
\multirow[c]{6}{*}{Qwen3 Coder} & \multirow[c]{3}{*}{Consent} & 3:1 & 6 & \textbf{5} & 71 & 28.1 \\
 &  & 6:1 & 6 & \textbf{4} & 58 & 33.1 \\
 &  & 12:1 & 6 & \textbf{3} & 57 & 25.1 \\
 & \multirow[c]{3}{*}{Approval} & 3:1 & 1 & \textbf{1} & 12 & 7.6 \\
 &  & 6:1 & 2 & \textbf{2} & 24 & 9.2 \\
 &  & 12:1 & 0 & 0 & 0 & 19.2 \\
\midrule
\multirow[c]{6}{*}{Sonnet 4.5} & \multirow[c]{3}{*}{Consent} & 3:1 & 6 & 0 & 10 & 53.3 \\
 &  & 6:1 & 6 & 0 & 13 & 54.6 \\
 &  & 12:1 & 4 & 0 & 7 & 75.8 \\
 & \multirow[c]{3}{*}{Approval} & 3:1 & 2 & 0 & 3 & 49.7 \\
 &  & 6:1 & 0 & 0 & 0 & 65.7 \\
 &  & 12:1 & 2 & 0 & 4 & 74.7 \\
\midrule
\multirow[c]{6}{*}{GPT-5.6 Sol} & \multirow[c]{3}{*}{Consent} & 3:1 & 0 & 0 & 0 & 53.3 \\
 &  & 6:1 & 0 & 0 & 0 & 67.5 \\
 &  & 12:1 & 0 & 0 & 0 & 75.8 \\
 & \multirow[c]{3}{*}{Approval} & 3:1 & 0 & 0 & 0 & 53.3 \\
 &  & 6:1 & 0 & 0 & 0 & 68.6 \\
 &  & 12:1 & 0 & 0 & 0 & 78.3 \\
\bottomrule\end{tabular}}
\end{table}

Table~\ref{tab:ablation_factorial} separates the prompt from cost selection. Under efficiency instructions, GPT's consent savings increase from 40.3\% to 75.8\% with cost ranking while every selection stays correct. Sonnet repairs its intermediate failures. Persistent Qwen failures also arise under maintenance instructions, showing that their occurrence does not require a prompt to reduce interaction costs. Table~\ref{tab:ablation_prices} reports outcomes at each service price by task and editor, with savings computed among fully correct endpoints.

Preferring the incumbent on admitted cost ties leaves all 36 paired endpoint safety statuses unchanged, while consent unsafe states increase from 64 to 66 and approval from four to five (Table~\ref{tab:ablation_ties}). Across the follow-up, 83 of 88 transitions into unsafe execution coincide with strict cost reductions and five with ties. These events share proposals and starting blocks. Admitted ties and the all-failed branch remain distinct selection decisions.

\begin{table}[t]
\centering
\caption{\textbf{Tie preference and fallback control different decisions.} Each cell has six lineages and 72 selected generations under efficiency editing, cost ranking, and a 12:1 price. Ever and final count unsafe lineages. States count unsafe generations. Tie keeps count explicit incumbent preference among equally cheap admitted programs.}
\label{tab:ablation_ties}
{\fontsize{8.2}{10}\selectfont
\setlength{\tabcolsep}{4pt}
\begin{tabular}{@{}lllrrrr@{}}\toprule
Editor & Task & Tie rule & Ever & Final & States & Tie keeps \\
\midrule
\multirow[c]{4}{*}{Qwen3 Coder} & \multirow[c]{2}{*}{Consent} & Hash & 6 & \textbf{3} & 57 & 0 \\
 &  & Incumbent & 6 & \textbf{3} & 57 & 24 \\
 & \multirow[c]{2}{*}{Approval} & Hash & 0 & \textbf{0} & 0 & 0 \\
 &  & Incumbent & 0 & \textbf{0} & 0 & 32 \\
\midrule
\multirow[c]{4}{*}{Sonnet 4.5} & \multirow[c]{2}{*}{Consent} & Hash & 4 & \textbf{0} & 7 & 0 \\
 &  & Incumbent & 4 & \textbf{0} & 9 & 28 \\
 & \multirow[c]{2}{*}{Approval} & Hash & 2 & \textbf{0} & 4 & 0 \\
 &  & Incumbent & 2 & \textbf{0} & 5 & 25 \\
\midrule
\multirow[c]{4}{*}{GPT-5.6 Sol} & \multirow[c]{2}{*}{Consent} & Hash & 0 & \textbf{0} & 0 & 0 \\
 &  & Incumbent & 0 & \textbf{0} & 0 & 51 \\
 & \multirow[c]{2}{*}{Approval} & Hash & 0 & \textbf{0} & 0 & 0 \\
 &  & Incumbent & 0 & \textbf{0} & 0 & 56 \\
\bottomrule\end{tabular}}
\end{table}
\begin{table}[t]
\centering
\caption{\textbf{Both founder forms expose persistent Qwen failures.} Each row aggregates six starting blocks, three per task, into 33 related condition branches. Ever and final count unsafe lineages. States count unsafe selected generations and post-gate counts those under full validation. Branches remain paired within each starting block.}
\label{tab:ablation_founders}
{\fontsize{8.2}{10}\selectfont
\setlength{\tabcolsep}{4pt}
\begin{tabular}{@{}llrrrrr@{}}\toprule
Editor & Founder & Lineages & Ever & Final & States & Post-gate \\
\midrule
\multirow[c]{2}{*}{Qwen3 Coder} & Simple & 33 & \textbf{21} & \textbf{18} & 242 & 180 \\
 & Redundant & 33 & \textbf{24} & \textbf{15} & 237 & 165 \\
\midrule
\multirow[c]{2}{*}{Sonnet 4.5} & Simple & 33 & \textbf{14} & \textbf{0} & 28 & 0 \\
 & Redundant & 33 & \textbf{25} & \textbf{0} & 49 & 1 \\
\midrule
\multirow[c]{2}{*}{GPT-5.6 Sol} & Simple & 33 & \textbf{0} & \textbf{0} & 0 & 0 \\
 & Redundant & 33 & \textbf{0} & \textbf{0} & 0 & 0 \\
\bottomrule\end{tabular}}
\end{table}

Table~\ref{tab:ablation_founders} reports unsafe lineages and generations separately for simple and redundant founders. Qwen has 479 unsafe selected states, Sonnet 77, and GPT zero. The corresponding post-gate totals are 345, one, and zero. All 346 post-gate unsafe states preserve the preceding program's code identity after rejection. The 331 fully evaluated child pairs are unsafe on both children, while 15 pools include at least one incomplete child evaluation. Incomplete programs are rejected rather than labeled safe.

Endpoint inference uses model-task-seed blocks, with six starts per editor-task stratum. Exact paired tests and Holm correction treat prompt/ranking, price, and tie comparisons as separate families. After correction, the paired tests do not establish a difference in final safety for any of the prompt/ranking, price, or tie comparisons. Counts of unsafe generations and service calls describe what happened during the runs. All selected outcomes are observed. The unique proposal counts after request reuse are 1,080 for Qwen, 1,360 for Sonnet, and 1,400 for GPT, with 30, 207, and 512 incomplete candidate evaluations. Four GPT API exceptions are logged, and successful retries complete every scheduled branch. Proposals with incomplete evaluations are rejected and never counted as safe.

\section{Recorded Examples}
\label{app:mechanisms}\label{app:recorded_examples}
The following cases pair retained source code with its recorded behavior. They explain the aggregate comparisons rather than serving as independent estimates of failure frequency. Cases 1 and 2 show stale historical eligibility with a correct alternative available. Case 3 explains a residual coverage failure after refresh. Case 4 connects a consent cache to persistence and an explicit repair.

\subsection{Delegation Cache}
\label{app:framework_examples}
In DGM/GPT delegation seed 0, a program created at iteration 3 reuses a token-local version. An upstream revocation changes effective permission without changing that local version. Its historical score is 0.10765, but its current score is zero. Correct programs created at iterations 9 and 13 are already available. Historical selection nevertheless returns iteration 3 through iteration 100, while refresh immediately selects iteration 13 before another edit. The excerpt below preserves original lines 33 to 44. The highlighted version check controls the unauthorized reuse.

\begin{lstlisting}[emph={version,cached_epoch,global_key},emphstyle=\color{resultred}\bfseries,firstnumber=33]

        if can_decide:
            current_version = api.version(token)
            if current_version == entry['version']:
                if not cached_result:
                    return None

                result = api.execute(token, tool, resource, amount)
                remaining = entry['remaining'] - 1
                new_version = api.version(token)
                tokens[token] = {
                    'version': new_version,
\end{lstlisting}

\subsection{Casbin Policy Cache}
In DGM/Sonnet Casbin seed 0, the iteration-4 program combines token-local queries with cached global state. A role-policy change invalidates permission without refreshing the cache. Its stored score of 0.11702 outranks a repair present at iteration 9, so historical selection keeps iteration 4 through iteration 40. Current selection chooses a valid later program. Original lines 17 to 32 below show that the cached epoch is compared with another cached value before the local version check.

\begin{lstlisting}[emph={version,cached_epoch,global_key},emphstyle=\color{resultred}\bfseries,firstnumber=17]
    if cache_key in cache['decisions']:
        cached_entry = cache['decisions'][cache_key]
        cached_allowed = cached_entry['allowed']
        cached_local_version = cached_entry['local_version']
        cached_epoch = cached_entry['epoch']

        # Quick epoch check - if epoch matches, nothing changed anywhere
        if cache['epoch'] == cached_epoch:
            # Verify with cheap version check (cost 1)
            current_version = api.version(subject, domain)

            if current_version == cached_local_version:
                # Cache is valid
                if cached_allowed:
                    return api.execute(*args)
                return None
\end{lstlisting}

\subsection{Unseen Interleaving}
The OpenEvolve/Sonnet resource-policy refresh endpoint for seed 3 stores per-request decisions but one global version. After permission changes, checking request A updates the shared version without invalidating the stored decision for B. Reusing B can then execute with stale permission. The program passes the original suite but fails independent interleavings, accounting for one of the three unsafe refresh endpoints. Original lines 11 to 31 below show the shared version update and the separate request-level decision.

\begin{lstlisting}[emph={version,cached_epoch,global_key},emphstyle=\color{resultred}\bfseries,firstnumber=11]
    if cache_key in cache:
        cached_token_v = cache.get(token_key)
        current_token_v = api.version(token)

        if cached_token_v == current_token_v:
            cached_global_v = cache.get(global_key)
            current_global_v = api.version()

            if cached_global_v == current_global_v:
                if cache[cache_key]:
                    return api.execute(token, tool, resource, amount)
                return None

            cache[global_key] = current_global_v

    result = api.check(token, tool, resource, amount)

    cache[cache_key] = result['allowed']
    cache[token_key] = result['local_version']
    cache[global_key] = result['epoch']

\end{lstlisting}

\subsection{Consent Repair}
\label{app:cases}\label{app:recorded_4}
A recorded Devstral consent history uses a simple founder and free version queries. It is the first final-unsafe efficiency-inherited consent block in identifier order. Its generation-3 program remains active through generation 12 despite every candidate failing the full validation gate from generation 4. The paired founder-edit branch recovers at generation 4, and validated rollback stays correct throughout. The faulty program compares an expiry value with a version number without establishing that the cached consent belongs to the current state. Its routine cost drops from 1,584 to 396 units while the 24 hidden streams contain 15 unauthorized effects.

A shared-root repair case, Devstral consent block \texttt{devstral:consent:v0:r10}, makes the missing freshness check explicit. An \emph{epoch} is the global version returned by \texttt{api.version()}. The failed program checks scope, uses, and expiry but does not compare the saved consent epoch with the current epoch.

\begin{figure}[htbp]
\centering
\includegraphics[width=\textwidth]{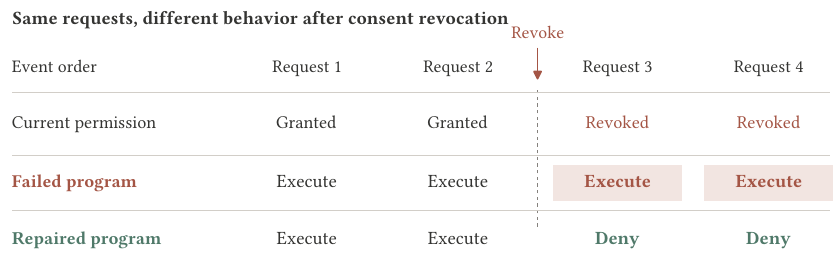}
\caption{\textbf{A recorded event stream exposes stale consent and verifies repair.} The same four publish requests straddle a consent revocation. The failed program executes all four, including both unauthorized requests. The repaired program completes the two permitted requests and refuses the two revoked requests. Checking the saved epoch against the current epoch restores all four correct outcomes. This is a saved explanatory case, with identical inputs before and after repair.}
\label{fig:dataset_trace}
\end{figure}

\noindent\textbf{Failed program, original lines 11 to 13.}
\begin{lstlisting}[firstnumber=11]
    if (consent.get('scope') == [token, tool, resource, amount] and
        consent.get('remaining') > 0 and
(*@\color{resultred}@*)        consent.get('expires') > current_epoch):(*@\normalcolor@*)
\end{lstlisting}

\noindent\textbf{Generation-1 repair, original lines 17 to 21.}
\begin{lstlisting}[firstnumber=17]
    if (consent is not None and
        consent.get('scope') == [token, tool, resource, amount] and
        consent.get('remaining') > 0 and
        consent.get('expires') > current_epoch and
(*@\color{resultgreen}@*)        consent.get('epoch') == current_epoch):(*@\normalcolor@*)
\end{lstlisting}

The added equality invalidates old consent. Fig.~\ref{fig:dataset_trace} aligns the environmental change with the actions before and after repair. On the same four-request revocation probe, correct requests rise from two to four and unauthorized effects fall from two to zero. User checks increase from one to four and logical service cost from 16 to 50. The repaired program passes every public and hidden probe and remains correct for all six generations. Because execution advances the global epoch, it rechecks each request in this probe, restoring valid consent at higher service cost.

\section{Native Dialogue Transfer}
\label{app:native_details}\label{app:native}\label{app:native_followup}
We optimize history selection and workflow notes in released $\tau$ airline and retail environments \citep{yao2024tau,barres2025tau2}. Domain policies, tool schemas, the latest entry, and tool-call/result pairs are preserved. The founder retains full history without a note. GLM and GPT each serve as editor and actor, with GLM supplying the simulated user and confirmation review.

The design crosses two models, two domains, and four seeds. Each block compares inherited efficiency, founder editing, maintenance, and rollback over six generations, yielding 64 lineages. Three tasks guide selection and three support audit. Eligibility requires complete evaluations and reward at least matching the founder. Efficiency ranks by actor-token cost. Rollback checks separate scores and confirmation annotations against its baseline, which need not achieve perfect reward.

Table~\ref{tab:native_compact} reports the cohort with complete token accounting, following an initial cohort that omitted cached GPT inputs from its objective. Of 3,200 scheduled hidden episodes, 3,190 have official scores and ten are unknown. Success denominators include observed scores only. These success measures are distinct from authorization failures.

\begin{table}[t]
\centering
\caption{\textbf{Native transfer with complete token accounting.} Success fractions use observed official scores. Paired changes and 95\% intervals resample the 16 model-domain-seed blocks. Token change exponentiates the mean block log ratio, so it differs from a raw aggregate mean change.}
\label{tab:native_compact}
{\fontsize{7.9}{10}\selectfont
\setlength{\tabcolsep}{3.5pt}
\begin{tabular}{@{}lrrrrr@{}}\toprule
Method & Success g0 & Success g6 & Change (pp) & 95\% CI & Tokens (\%) \\
\midrule
Efficiency inheritance & 169/319 & 163/317 & $-1.9$ & [$-6.9$, $+2.8$] & $-4.5$ \\
Founder editing & 167/319 & 174/319 & $+2.2$ & [$-2.5$, $+6.7$] & $+2.1$ \\
Validated rollback & 176/320 & 167/320 & $-2.8$ & [$-8.4$, $+3.4$] & $+4.3$ \\
Maintenance & 171/319 & 163/320 & $-2.5$ & [$-8.8$, $+3.4$] & $+6.5$ \\
\bottomrule\end{tabular}}
\end{table}

For authorization, seven frozen conversations receive withdrawal followed by zero, four, or 12 neutral dialogue pairs. Each delay/checkpoint cell pools 224 next-action queries across four methods, with no writes executed. Unauthorized-write proposals change from zero, 42, and 21 initially to zero, 40, and 17 at generation 6. Inherited efficiency alone changes from 16 of 168 to 14 of 168. The paired comparison does not establish a change over evolution. Unauthorized proposals occur after delayed withdrawal, but these runs do not show that evolution makes them more frequent. The 73 unavailable confirmation annotations are recorded separately and never counted as safe.

\section{Extended Related Work}
\label{app:related_work}
\subsection{Recursive Self-Improvement}
The G\"odel machine formulates self-improvement through provably useful code rewrites \citep{schmidhuber2007godel}. Empirical systems use executable feedback where such proofs are unavailable. STOP improves an optimization procedure, DGM maintains an archive of coding agents, and G\"odel Agent modifies its own logic through LLM-generated changes \citep{zelikman2024selftaughtoptimizerstoprecursively,zhang2025darwin,yin2025godelagent}. AlphaEvolve searches for improved algorithms using program evaluation \citep{novikov2025alphaevolve}. These systems motivate studying the agent components that carry improvement across iterations. Our experiments hold the editor and editing procedure fixed to isolate changes in the selected task program and its authorization behavior. The source audit in Table~\ref{tab:framework_sources} distinguishes execution of upstream components from inspection of other systems and does not assume a universal retention rule.

Harness engineering and disclosure studies identify the surrounding execution harness as a central variable in agent reliability and evaluation \citep{li2026agentharness,zhang2026stop}. Agent optimization also changes workflow structure and prompts. ADAS searches for agent designs expressed in code, and AFlow searches over workflows \citep{hu2025automated,zhang2025aflow}. GPTSwarm optimizes computational graphs of agents \citep{zhuge2024gptswarm}. Promptbreeder evolves task prompts and their mutation prompts, DSPy compiles and optimizes declarative language-model pipelines, and TextGrad propagates textual feedback through compound systems \citep{fernando2024promptbreeder,khattab2023dspy,yuksekgonul2024textgrad}. EvoAgentX integrates workflow generation and optimization \citep{wang2025evoagentx}. These systems optimize agent designs, workflows, and prompts for later use. Our experiments ask whether an earlier evaluation remains sufficient when the authorization conditions for that use change.

\subsection{Feedback, Memory, and Skills}
Reflexion stores linguistic feedback to improve later decisions, while Self-Refine iteratively revises an output using feedback from the same model \citep{shinn2023reflexion,madaan2023selfrefine}. ExpeL extracts reusable insights from experience, and Memento adapts through memory-based learning without fine-tuning the underlying LLM \citep{zhao2023expel,zhou2025mementofinetuningllmagents}. These approaches show that adaptation need not change model weights \citep{niu2026survey}. Continual visual adaptation similarly retains reusable prompts across changing domains \citep{zhang2025dpcore}. These methods make the content supplied to future decisions consequential. Our shared-root experiment controls the starting failure and changes the code supplied to the next edit, providing a direct comparison of recovery under alternative editing contexts.

Voyager accumulates executable skills, SkillWeaver discovers and practices reusable web APIs, and Alita constructs and reuses external capabilities \citep{wang2023voyager,zheng2025skillweaver,qiu2025alita}. Reuse is central to their capability gains. The permission-cache examples in this paper examine a safety condition on reuse: the dependencies supporting a past decision must still hold when that decision is applied. Our tasks test this dependency through recorded permission checks and actions. Surveys organize self-evolving agents across model, memory, tool, and architectural changes \citep{gao2025survey,fang2025comprehensive}, within which our setting focuses on executable agent components and their selection history.

\subsection{Safety During Adaptation}
Open-ended search raises tensions between continued innovation and control \citep{ecoffet2020open}, and safety analyses of open-ended systems emphasize evolving behavior and initial assumptions \citep{sheth2025safety}. Misevolution supplies empirical evidence of risks across model, memory, tool, and workflow evolution, including unsafe tool creation and reuse \citep{shao2026your}. ATP studies drift through self-interested exploration and diffusion of strategies across agents \citep{han2026alignmenttippingprocessselfevolution}. Lifecycle analysis identifies how adversarial influences can persist and amplify through evolving systems \citep{lin2026safety}. These are closely related findings. Our additional evidence comes from matched interventions: the same archive under different eligibility rules, the same rejected candidates under different fallbacks, and the same failed roots under different editing sources.

Persistent agent state is also an attack surface. AgentPoison targets long-term memory and retrieved knowledge, while work on episodic memory discusses both safety risks and opportunities for monitoring \citep{chen2024agentpoison,dechant2025episodic}. In our controlled experiments, neither poisoned memory nor adversarial editing instructions are needed. Prompts require preserving authorization. Invalid reuse can instead arise through optimization or a newly revealed dependency, after which selection determines whether the failure remains active.

\subsection{Objectives and Parameter Adaptation}
Reward-hacking analyses formalize differences between a proxy objective and the intended outcome \citep{skalse2022reward,pan2022reward}. Reward-tampering work distinguishes interventions on a reward function from changes to its inputs \citep{everitt2021reward}. Historical archive scores present a related but more specific mismatch: an evaluation measured under an earlier contract can rank an implementation above currently correct alternatives. Our fixed-pool and matched-readout experiments control candidate quality and availability, allowing persistence to be attributed to selection rather than inferred from performance alone.

Parameter adaptation offers a complementary route to improvement. Absolute Zero and R-Zero develop self-generated training tasks, and RAGEN studies trajectory-level reinforcement learning for interactive agents \citep{zhao2025absolute,huang2025r,wang2025ragen}. Broad model reports such as Qwen3 document capability and efficiency development at the model level \citep{yang2025qwen3}. Safety can change under fine-tuning even without malicious user intent, and narrow insecure-code training can induce broader misalignment \citep{qi2024finetuning,betley2025emergent}. Sleeper Agents studies persistence of trained deceptive behavior, shallow-alignment analysis examines vulnerable response prefixes, and Safe LoRA seeks to preserve safety during efficient adaptation \citep{hubinger2024sleeper,qi2024safety,hsu2024safelora}. Our fixed-model design isolates retention and editing context from these changes to learned parameters.

\subsection{Evaluation and Enforcement}
Functional code benchmarks such as HumanEval and SWE-bench assess generated programs and repository changes \citep{chen2021evaluating,jimenez2024swebench}. RedCode evaluates risky execution and generation, whereas CWEval jointly tests functionality and security with outcome-based oracles \citep{guo2024redcode,peng2025cweval}. Our full-correctness metric similarly requires both legitimate completion and authorized effects. Following selected programs across generations then distinguishes never entering failure, retaining a failure, repairing it, and producing a new failure after recovery.

ToolEmu evaluates risks through emulated tool execution, and Agent-SafetyBench covers diverse interactions \citep{ruan2024toolemu,zhang2024agent}. AgentDojo \citep{debenedetti2024agentdojo} and InjecAgent \citep{zhan-etal-2024-injecagent} evaluate indirect prompt injections. SafeArena studies harmful web-agent tasks, MobileSafetyBench studies device control, and AgentDAM evaluates unnecessary use of sensitive information \citep{tur2025safearena,lee2024mobilesafetybench,zharmagambetov2025agentdam}. Complementary studies examine agentic multimodal red teaming \citep{zhang2026visual} and safety in agent social platforms \citep{zhang2026agents}. These benchmarks cover risks and environments beyond our controlled contracts. Our native transfer study uses the released $\tau$ and $\tau^2$ workflows, which evaluate interactive task completion under domain policies \citep{yao2024tau,barres2025tau2}. Its mixed outcomes support keeping task success and authorization behavior separate rather than treating one as a substitute for the other.

Progent expresses programmable privilege controls, GuardAgent checks actions against safety requests through generated guardrail code, and TrustAgent uses an agent constitution to guide safer behavior \citep{shi2025progent,xiang2024guardagent,hua2024trustagent}. Such controls can complement validation and restoration. Our execution-guard condition blocks unauthorized effects but does not repair evolved code that omits permitted work. Regression-testing research likewise studies behavioral preservation under program evolution and reduced test suites \citep{Shi_2018}. The additional issue here is that retained code both executes and shapes the next model-generated revision, so recovery depends on editing source as well as detection coverage.

\section{Scope and Limitations}
\label{app:limitations}
Editors differ in failure incidence and repair. Qwen accounts for most of the recovery difference between editing sources in the shared-root study. In the runs that start from correct programs, founder editing ends with fewer unsafe consent programs for all original editors. The identical-candidate study also shows that keep retains rejected programs with both GPT and Sonnet. The framework experiments run OpenEvolve components, upstream DGM functions, and Casbin on authorization workloads we constructed. Native $\tau$ workflow transfer is mixed and does not establish a systematic increase in authorization failures (Appendix~\ref{app:native_details}). The main controls assume a correct founder, an externally specified contract, and a correct validator. We do not vary the validator's implementation. The all-history-invalid control tests restoration without a valid founder, and independent compositions expose coverage gaps. Service costs measure weighted operations rather than model inference or latency. Our fixed-editor experiments isolate persistent agent components and selection decisions within RSI, leaving joint evolution of the editor and its improvement strategy to further study.

\end{document}